\documentclass[pra,aps,amsfonts,amsmath,superscriptaddress,onecolumn,showpacs,floatfix]{revtex4}
\usepackage{amsfonts}
\usepackage{amsmath}
\usepackage{bm}
\usepackage{epsfig}
\usepackage{mathrsfs}
\usepackage{graphics}

\usepackage[usenames]{color}
\usepackage{ulem}

\begin{document}

\title {\bf Second-order equation of state with the full Skyrme interaction: 
toward new effective interactions for beyond mean-field models}

\author{K. Moghrabi}
\address{Institut de Physique Nucl\'eaire,
 Universit\'e Paris-Sud, IN2P3-CNRS, F-91406 Orsay Cedex, France}
\author{M. Grasso}
\address{Institut de Physique Nucl\'eaire,
 Universit\'e Paris-Sud, IN2P3-CNRS, F-91406 Orsay Cedex, France}
\author{X. Roca-Maza}
\address{INFN, Sezione di Milano, Via Celoria 16, 20133 Milano, Italy}
\author{G. Col\`o}
\address{INFN, Sezione di Milano, Via Celoria 16, 20133 Milano, Italy}
\address{Dipartimento di Fisica, Universit\`a degli Studi di Milano, 
Via Celoria 16, 20133 Milano, Italy}


\begin{abstract}

In a quantum Fermi system the energy per particle calculated at the second order 
beyond the mean-field approximation diverges if a zero-range interaction is 
employed. We have previously analyzed this problem in symmetric nuclear matter 
by using a simplified nuclear Skyrme interaction, and proposed a strategy to treat 
such a divergence. In the present work, we extend the same strategy to the case of 
the full nuclear Skyrme interaction. Moreover we show that, in spite of the strong 
divergence ($\sim$ $\Lambda^5$, where $\Lambda$ is the momentum cutoff) related to the 
velocity-dependent terms of the interaction, the adopted cutoff regularization can 
be always simultaneously performed for both symmetric and nuclear matter with 
different neutron-to-proton ratio. This paves the way to applications to finite 
nuclei.

\end{abstract}

\vskip 0.5cm \pacs{21.60.Jz,02.60.Ed,21.30.-x,21.65.Mn} \maketitle


\section{Introduction}
Mean-field theories are widely employed to analyze a large variety of many-body 
systems where an independent-particle picture can be adopted. In nuclear physics, 
mean-field approaches generally lead to satisfactory results when applied to 
several bulk properties of atomic nuclei such as masses, radii, or ground-state 
deformations. In some cases, however, in nuclear physics as well as in other domains 
of many-body physics, mean-field approaches need definite improvements. For 
instance, in the nuclear case, mean-field models do not predict accurately the 
single-particle spectra (namely, energies, spectroscopic factors and fragmentation 
of the single-particle states, especially close to the Fermi energy): 
the mean-field Hartree-Fock (HF) occupation numbers are strictly equal to 0 
or 1 and, therefore, the spectroscopic factors of single-particle states cannot 
be reproduced. To improve the theoretical description, one may introduce the 
coupling between nucleon individual degrees of freedom and collective degrees of 
freedom by means of 
the particle-vibration coupling approach \cite{bernard,colo,ring}. This model constitutes 
an example of beyond mean-field theories. 

A conceptual problem is often encountered when 
beyond mean-field theories are employed in 
nuclear physics where the most widely 
used interactions are phenomenological. That is, 
the parameters of these interactions already take into 
account in an effective way several 
correlations. Hence, their use in beyond mean-field models 
(where correlations are explicitly introduced) 
may lead to a double counting. 
This difficulty could be removed in principle 
by adjusting the parameters directly at the beyond mean-field level.  

An additional practical problem appears in 
beyond mean-field theories where the interaction 
has zero range. In the nuclear case, 
all the terms of the Skyrme interaction, and the 
density-dependent and spin-orbit terms of the Gogny 
interaction have zero range. Due to this, 
ultraviolet divergences are generated in 
the diagrams beyond HF,  
so that the numerical results are dependent on 
the choice of the energy cutoff.
 
Different procedures to regularize the divergences that appear beyond HF may be adopted. For example, in the context of the Hartree-Fock-Bogoliubov or Bogoliubov-de Gennes model (where an ultraviolet divergence is generated if a zero-range pairing interaction is used) a currently employed regularization is based on the so-called pseudo-potential prescription \cite{huang}. This procedure has been proposed in nuclear \cite{bulgac} and in atomic \cite{bruun,grasso} physics. Another example is the cutoff regularization which is adopted in this work, where we change the cutoff and adjust simultaneously the parameters of the Skyrme effective interaction to keep the energy per particle
at the beyond mean-field approximation independent of the cutoff regulator. 
A third example is the elegant procedure of the dimensional regularization which has been developed in the 70s in the context of the electroweak theory \cite{dr1,dr2,dr3} and which is widely employed nowadays in perturbative quantum field theories to regularize the divergent integrals. The basic idea of this regularization technique consists in replacing the dimension $d$ of the integrals with a continuous variable. The dimensional regularization (at variance with the cutoff regularization) preserves translational, gauge and Lorentz symmetries \cite{leib} and this explains its wide diffusion and success in the framework of quantum field theories.   

The ultraviolet divergence appearing beyond the mean-field approach has been discussed in a previous 
work \cite{mog}, where the nature of the 
divergence has been analyzed in the case of 
symmetric nuclear matter within a simplified 
$t_0-t_3$ Skyrme model. The used interaction was $V(\vec{r}_1,\vec{r}_2 )=g(\rho)\delta (\vec{r}_1 - \vec{r}_2 )$, where the coupling constant 
is density dependent and equal to $g(\rho)=t_0 +\frac{t_3}{6}\rho^{\alpha}$; $t_0$, $t_3$ and $\alpha$ are parameters. The second-order 
correction to the mean-field equation of state (EoS), that is,  
 the lowest-order term beyond HF,  
has been studied analytically in detail. 
The precise nature of the divergence, which is 
linear in the momentum cutoff $\Lambda$,  
can be put in evidence by using the results of Ref. \cite{mog}. 
In such a case, the linear divergence in 
$\Lambda$ can be explicitly shown by taking the 
asymptotic expansion (for $\Lambda \rightarrow \infty$) of Eq. (8) of 
Ref. \cite{mog}. This expansion reads 

\begin{equation}
\frac{-11 + 2 \log{2}}{105}+\frac{\Lambda}{9k_F}-\frac{2 k_F}{45 \Lambda} + O(\frac{k_F^2}{\Lambda^2}).
\end{equation}

For different values of $\Lambda$, new 
sets of parameters ($t_0$, $t_3$ and $\alpha$) have been adjusted to 
reproduce a reference  
EoS for symmetric matter. In practice, the mean-field EoS
of the SkP Skyrme force \cite{skp} has been 
chosen as a benchmark. 
This choice has been suggested by the fact that, for the SkP 
parametrization, only the $t_0$ and $t_3$ terms actually 
appear in the EoS for symmetric matter.
In the framework of such cutoff regularization, 
the choice of $\Lambda$ has been dictated by 
physical criteria. As already mentioned 
in Ref. \cite{mog}, since in 
low-energy nuclear physics 
nucleons are treated as 
point 
particles, $\Lambda$ must be definitely smaller than 2 fm$^{-1}$.
This value of $\Lambda$ will be the maximum cutoff adopted in this work. 

In the present work, the approach of 
Ref. \cite{mog} is considerably extended.
The full Skyrme interaction is taken into account and both 
symmetric and asymmetric nuclear matter (including the
limiting case of pure neutron matter) are considered. 
The latter point is particularly important
since we show here that the obtained  
interaction fitted for several values of $\Lambda$  to a chosen benchmark EoS 
is very accurate for a wide range of densities and neutron-proton 
asymmetries as well as for large values of the momentum cutoff.
At variance with Ref. \cite{mog}, we have chosen here the SLy5 \cite{sly5} Skyrme parametrization to generate 
the mean-field EoS which has been adopted as a benchmark. 

The article is organized in the following way: in Sec. II 
the analytical expressions of the second-order contribution 
associated to the full Skyrme interaction
are described, and the nature of the 
divergence is analyzed. In Sec. III the 
numerical results for the EoS (with different values 
of $\Lambda$) are displayed, and 
the fit of the 9 Skyrme parameters is firstly done separately in the 
case of symmetric (Subsec. III.A), pure neutron (Subsec. III.B), 
and asymmetric (Subsec. IV.C) matter. Finally, we show
a global fit for the three cases 
in Subsec. III.D.  
Conclusions are drawn in Sec. IV. 

\section{Second-order equation of state with the Skyrme interaction}

As done in Ref. \cite{mog}, we 
treat the EoS of nuclear matter by adding 
the second-order correction to the first-order mean-field energy. 
The EoS $E/A$ is thus written as the sum of four diagrams 
which are displayed in Fig. 1 and which represent the direct (left) and exchange (right) first-order (upper line) and second-order (lower line) contributions. 
As mentioned in Sec. I, we use in this paper a
standard Skyrme interaction like SLy5 in its complete form, namely

\begin{eqnarray}
V({\vec r}_1, {\vec r}_2) &=& t_0(1+x_0P_{\sigma}) \delta({\vec r}_1-
{\vec r}_2)
+ \frac{1}{2} t_1 (1+x_1P_{\sigma})[{\vec k}'^{2}
\delta({\vec r}_1-{\vec r}_2)+
\delta({\vec r}_1-{\vec r}_2){\vec k}^{2}] \nonumber \\
&+& t_2(1+x_2P_{\sigma}){\vec k}'\cdot\delta({\vec r}_1-
{\vec r}_2){\vec k} +
\frac{1}{6} t_3 (1+x_3P_{\sigma})\rho^{\alpha}({\vec R})
\delta({\vec r}_1-{\vec r}_2)\nonumber\\
&+& iW_0 ({\sigma}_1+{\sigma}_2)\cdot[{\vec k}'\times
\delta({\vec r}_1-{\vec r}_2){\vec k}]~,
\label{eq2.1-1}
\end{eqnarray}
where ${\vec R} =\frac{1}{2} ({\vec r}_1+{\vec r}_2)$,
${\vec k} = \frac{1}{2i}({\vec \nabla}_1-{\vec \nabla}_2)$, 
$\vec k'$ is the hermitian conjugate
of $\vec k$ (acting on the left), 
$P_{\sigma}=\frac{1}{2}(1+{\sigma}_1\cdot{\sigma}_2)$
is the spin-exchange operator. The ten parameters 
$t_i$, $x_i$, $\alpha$ and $W_0$ characterize a given
Skyrme set. In uniform matter the spin-orbit force
does not play any role, and the associated parameter
$W_0$ does not show up in any of the quantities discussed
below. Consequently, we consider only the remaining nine free
parameters. 

Using the general Skyrme force of Eq. (\ref{eq2.1-1}), 
the energy per particle in uniform matter can be easily
written. The mean-field (or HF) result can be
found in several papers. The result including both
HF and the second-order correction reads 
\begin{widetext}
\begin{eqnarray}
\nonumber
\frac{E}{A}(\delta,\rho,\Lambda) &=& 
\frac{3\hbar^2}{10m}\left(\frac{3\pi^2}{2}\rho \right)^{\frac{2}{3}} G_{5/3} + 
\frac{1}{8} t_0 \rho [2(2+x_0)-(1+2x_0)G_2] 
+ \frac{1}{48} t_3 \rho^{\alpha+1} [2(2+x_3)-(1+2x_3)G_2] \\ &+& 
\frac{3}{40}\left(\frac{3\pi^2}{2}\right)^{\frac{5}{3}}
\rho^{\frac{5}{3}}\left[ \Theta_v G_{5/3} + \frac{1}{2} (\Theta_s -2 \Theta_v) 
G_{8/3} \right] + \frac{\Delta E^{(2)}(\delta, \rho, \Lambda)}{A}, 
\label{eq15}
\end{eqnarray}
\end{widetext}
where $\delta$ is the 
asymmetry parameter, 
\begin{equation}
\delta=\frac{\rho_n - \rho_p}{\rho_n+\rho_p}.
\end{equation}
In Eq. (4) $\rho_n$ and $\rho_p$ are equal to the neutron and proton densities, respectively; $\delta=0$ and $\delta=1$ represent the extreme 
cases of symmetric and pure neutron matter, respectively.
The following notation is used in Eq. (3), 
\begin{eqnarray}
\nonumber
G_{\beta} = \frac{1}{2} [(1+\delta)^{\beta}+(1-\delta)^{\beta}], \\
\nonumber 
\Theta_s = 3t_1 + t_2 (5+4x_2), \\
\nonumber
\Theta_v = t_1 (2+x_1)+t_2(2+x_2).
\end{eqnarray}

We observe that the second-order term, which is the last
term in Eq. (\ref{eq15}), depends not only on 
$\rho$ and $\delta$ but on the momentum cutoff $\Lambda$ as well. 
Its derivation is discussed in what follows. 

To have a more compact notation, let us write the second-order correction as $\Delta E^{(2)}$ by omitting the explicit dependence on $\delta$, $\rho$ and $\Lambda$, 
\begin{equation}
\Delta E^{(2)}=(\Delta E + \Delta E^x)_{nn}+(\Delta E + \Delta E^x)_{pp}+\Delta E_{np}.
\label{eq1}
\end{equation}

The direct $\Delta E$ and the exchange $\Delta E^x$ contributions for the neutron-neutron ($nn$) and proton-proton ($pp$) channels are written in a box of volume $\Omega$ as 

\begin{widetext}
\begin{eqnarray}
\Delta E_{ii} = \frac{1}{2} \frac{\Omega^3}{(2\pi)^9} \int_{C_I} d^3 \vec{k}_1\;d^3 \vec{k}_2\;d^3 \vec{q}
\;\frac{v^2(\vec{q})}{\epsilon_{k_1}+\epsilon_{k_2}-\epsilon_{k_1+q}-\epsilon_{k_2-q}},
\\
\Delta E_{ii}^x = -\frac{1}{2} \frac{\Omega^3}{(2\pi)^9} 
 \int_{C_I} d^3 \vec{k}_1\;d^3 \vec{k}_2\;d^3 \vec{q}\;  
\frac{v(\vec{q}) v(\vec{k}_1-\vec{k}_2+\vec{q})}
{\epsilon_{k_1}+\epsilon_{k_2}-\epsilon_{k_1+q}-\epsilon_{k_2-q}},  
\label{eq3}
\end{eqnarray}
\end{widetext}
where $v$ represents the interaction in momentum space and 
$i$ denotes either $p$ or $n$. To simplify the notation, 
in $v^2(\vec{q})$ and 
$v(\vec{q})v(\vec{k}_1-\vec{k}_2+\vec{q})$ we have
included the factors resulting from the fact that the
sum over spin and isospin indices has been performed.
So these quantities read
\begin{eqnarray}
\nonumber
v^2(\vec{q}) &=& 4(\gamma_1 + \gamma_2)+(8\gamma_3+4\gamma_4) q^2 + 4(\gamma_5+\gamma_6) q^4, \\
v(\vec{q}) v(\vec{k}_1-\vec{k}_2+\vec{q}) 
&=& (2\gamma_1+8\gamma_2)+(2\gamma_3+4\gamma_4)\left[q^2+(\vec{k_1}-\vec{k_2}+\vec{q})^2\right]+(8\gamma_5+2\gamma_6)
\;q^2(\vec{k_1}-\vec{k_2}+\vec{q})^2.
\end{eqnarray}
The parameters $\gamma_i$ in Eq. (8) are listed in the following table,
\begin{center}
\begin{tabular}{ccccccc}
\hline
$\gamma_1$ \;\;\;\;&\;\;\;\; $\gamma_2$ \;\;\;\;&\;\;\;\; $\gamma_3$ \;\;\;\;&\;\;\;\; $\gamma_4$ 
\;\;\;\;&\;\;\;\; $\gamma_5$ \;\;\;\;&\;\;\;\; 
$\gamma_6$ \\
\hline
$ t^{2}_{03}+x^{2}_{03}$ \;\;\;\;&\;\;\;\;  $t_{03}\;x_{03}$ \;\;\;\;&\;\;\;\;  $t_{03}t_{12}+x_{03}x_{12} $ 
\;\;\;\;&\;\;\;\;  $t_{03}x_{12}+x_{03}t_{12} $ \;\;\;\;&\;\;\;\;  $t_{12}\;x_{12}$ \;\;\;\;&\;\;\;\; 
 $ t^{2}_{12}+x^{2}_{12}$ \\
\hline
\end{tabular}
\end{center} 
where the following notation has been adopted,
\begin{eqnarray}
\nonumber
t_{03} &=& t_0 + \frac{t_3}{6} \rho^{\alpha}, \\
\nonumber
x_{03} &=& t_0 x_0 + \frac{t_3 x_3}{6} \rho^{\alpha}, \\
\nonumber
t_{12} &=& t_1 + t_2, \\
\nonumber 
x_{12} &=& t_1 x_1 + t_2 x_2. 
\end{eqnarray}
In Eqs. (6) and (7) the energies $\epsilon$ are expressed as $\epsilon_{k}=\frac{\hbar^2k^2}{2m^*_{n,p}}$,
where $m^*_{n,p}/m$ is the effective mass for neutrons or protons that we have taken equal to its mean-field value \cite{sly5}, 
\begin{eqnarray}
\nonumber
\frac{m}{m^*_{n,p}}&=&1+\frac{1}{4} \frac{m}{\hbar^2} (\rho_n + \rho_p) \Theta_v +\frac{1}{4} \frac{m}{\hbar^2} \rho_{n,p}(\Theta_s - 2 \Theta_v) \\
&=& 1+\frac{1}{4} \frac{m}{\hbar^2} \rho \Theta_v +\frac{1}{4} \frac{m}{\hbar^2} \frac{\rho(1+\omega_{n,p}\delta)}{2}(\Theta_s - 2 \Theta_v),
\end{eqnarray}
where $\omega_n=1$ and $\omega_p=-1$. 
In what follows we shall also use the mean-field 
isoscalar effective mass $m_s^*/m$, namely
\begin{equation}
\left( \frac{m_s^*}{m} \right)^{-1} = 1+ \frac{1}{8} \frac{m}{\hbar^2}\Theta_s \rho,   
\end{equation}
and the parameters
\begin{equation}
\nonumber
b=\frac{m^*_s}{m^*_n}, \;\;\;\;\;\; c=\frac{m^*_s}{m^*_p}. 
\end{equation}

To clarify our compact notation, we stress that 
in the $nn$ and $pp$ channels, 
the $k_1$ and $k_2$ momenta appearing in the integrals of Eqs. (6) and (7) refer either both to protons or both to neutrons. The integration domain $C_I$ is given by 
$C_I=\left[|\vec{k}_1|<k_F;|\vec{k}_2| < k_F; |\vec{k}_1+\vec{q}|>k_F;|\vec{k}_2-\vec{q}| > k_F \right]$. 
This means that $\vec{k}_1$ and $\vec{k}_2$ represent hole states whereas $\vec{k}_1+\vec{q}$ and $\vec{k}_2-\vec{q}$  represent particle states. The Fermi momentum $k_F$ 
in $C_I$
refers to  the proton (neutron) Fermi momentum if $k_1$ and $k_2$ represent both protons (neutrons).

For the neutron-proton channel one has
\begin{widetext}
\begin{equation}
\Delta E_{np} = \frac{\Omega^3}{(2\pi)^9}  \int_{C_I} d^3 \vec{k}_1\;d^3 \vec{k}_2\;d^3 \vec{q}
\frac{2 v^2(\vec{q})}{\epsilon_{k_1}+\epsilon_{k_2}-\epsilon_{k_1+q}-\epsilon_{k_2-q}},
\label{eq4}
\end{equation}
\end{widetext}
where $v^2(\vec{q})$ is explicitly written in Eq. (8). The factor 2 in Eq. (11) comes from the sum of the $np$ and $pn$ channels. The momentum $k_1$ 
($k_2$) is associated to a neutron (proton). That means that,   
in the integration domain $C_I$ that we write formally in the same way as for the $nn$ and $pp$ cases, the $k_F$ associated to the $k_i$ of the neutron (proton) represents the neutron (proton) Fermi momentum.  
We have not specified so far the two different $k_F$ values in the equations to avoid a heavy notation. We will denote later 
the two Fermi momenta as $k_n$ and $k_p$ where, 

\begin{eqnarray}
k_n &=& \left( \frac{3}{2}\pi^2 \rho_n \right)^{1/3} = 
\left[ \frac{3}{4} \pi^2 \rho (1 + \delta) \right] ^{1/3}, \\
k_p  &=& \left( \frac{3}{2}\pi^2 \rho_p \right)^{1/3}= 
\left[ \frac{3}{4} \pi^2 \rho (1 - \delta ) \right] ^{1/3}.
\end{eqnarray}

We introduce the parameter $a$ depending on $k_n$ and $k_p$, 

\begin{equation}
a=\frac{k_p}{k_n}= \left(\frac{1-\delta}{1+\delta}\right)^{1/3} \le 1. 
\label{eq8}
\end{equation}

In general, as already mentioned, one can show that all the corrective terms 
$\Delta E$ are functions of the density $\rho$, of the asymmetry parameter 
$\delta$ and of the momentum cutoff $\Lambda$. Indeed, 
after some manipulations, the $pp$ and $nn$ contributions can be written as the sum of ten terms,  

\begin{equation}
\sum_{i=n,p}\frac{\Delta E_{ii}+\Delta E^x_{ii}}{A}(\delta, \rho,\Lambda)= \sum_{j=1}^{10} \chi_j(\delta, \rho)I_j(\delta,\rho,\Lambda),
\label{eq5}
\end{equation}
where the first 5 terms describe the $nn$ channel and the last 5 terms describe the $pp$ channel. Let us start with the $nn$ case.  
It can be seen that in the integrals $I_j$,  
\begin{widetext}
\begin{equation}
I_j(\delta,\rho,\Lambda)= \Gamma_j \int_0^{\frac{\Lambda}{2k_n}} du\;f_j(u)\;[\Theta(1-u)\;F^j_1(u) + 
\Theta(u-1)\;F^j_2(u)], \;\;\;\;\; j=1\ldots 5, 
\label{eq6}
\end{equation}
\end{widetext}
the $\rho$, $\delta$ and $\Lambda$ dependence enters in the upper limit of integration. 
The five coefficients $\Gamma_i$ and the five functions $f_i$ are listed in the following table:

\vspace{0.3cm}
\begin{center}
 \begin{tabular}{cccccccccc}
\hline
  $\Gamma_1$  ;\;&\;\;  $\Gamma_2$ \;\;&\;\;  $\Gamma_3$ \;\;&\;\;  $\Gamma_4$ \;\;&\;\;  $\Gamma_5$ \;\;&\;\; $f_1(u)$ \;\;&\;\; 
$f_2(u)$ \;\;&\;\; $f_3(u)$ \;\;&\;\; $f_4(u)$ \;\;&\;\; $f_5(u)$ \\
\hline
1/15 \;\;&\;\;  1/15 \;\;&\;\;  -1 \;\;&\;\;  4/15 \;\;&\;\;  1/15 \;\;&\;\; $u$ \;\;&\;\; $u^2$ \;\;&\;\; $u^2$ \;\;&\;\; $u^4$ \;\;&\;\; $u^4$ \\
\hline
\end{tabular}
\end{center}
\vspace{0.3cm}

The expressions of the ten functions $F_1^j(u)$, $F_2^j(u)$ (with $j$ running from 1 to 5) are provided in Appendix A. 
The coefficients $\chi_j$ appearing in Eq. (15) (for $j=1\ldots 
5$) are written as follows 
\begin{eqnarray}
\nonumber
\chi_1(\delta,\rho) &=& 8 \pi^3 C_{\Delta E} m_n^*k_n^7(t_{03}-x_{03})^2, \\
\nonumber
\chi_2(\delta,\rho) &=& 32 \pi^3 C_{\Delta E} m_n^*k_n^9(t_{03} t_{12}+x_{03} x_{12}), \\
\chi_3(\delta,\rho) &=& 64 \pi^3 C_{\Delta E} m_n^*k_n^9(t_{03} x_{12} +x_{03} t_{12}), \\
\nonumber
\chi_4(\delta,\rho) &=& 64 \pi^3 C_{\Delta E} m_n^*k_n^{11} t_{12}x_{12}, \\
\nonumber
\chi_5(\delta,\rho) &=& 64 \pi^3 C_{\Delta E} m_n^*k_n^{11}(t_{12}^2+x_{12}^2), 
\label{eq7}
\end{eqnarray}
where the following notation has been introduced:
\begin{eqnarray}
C_{\Delta E} &=& - \frac{8}{(2\pi)^9} \frac{1}{\hbar^2 \rho}. 
\end{eqnarray} 

The integral $I_1$ in Eq. (15) has already been encountered 
in the $t_0-t_3$ model treated in Ref. \cite{mog}, whereas the additional four integrals $I_{2,3,4,5}$ appear when the full interaction is considered. The last five terms in Eq. (\ref{eq5}) ($pp$ channel) are equal to the first five terms with the replacements $m^*_n \rightarrow m^*_p$ and $k_n \rightarrow k_p$ in the coefficients $\chi$. 
The replacement $k_n \rightarrow k_p$ is also done in the upper limit of the integral in Eq. (16). 
The coefficients $\Gamma_i$ and the functions $f_i$ and $F^i_{1,2}$, with $i$ running from 6 to 10, are equal to those already written for $i$ running from 1 to 5.  

For the neutron-proton channel we divide the region of integration into three parts, $0 < |q| < 2k_p$, $2k_p < |q| < 2k_n$ and $|q|>2k_n$ and we use the parameter $a$, 
defined in Eq. (14).
One can derive the following expression, 
\begin{widetext}
\begin{equation}
\frac{\Delta E_{np}}{A}= \frac{16\pi^3 C_{\Delta E}k_n^7}{15} \frac{m_s^*}{(bc)^3} 
\left[\int_0^a du\;u\;v^2(2k_nu)F^{abc}_1(u) + \int_a^1 du\;u\;v^2(2k_nu)\;F^{abc}_3(u) + 
\int_1^{\frac{\Lambda}{2k_n}} du\;u\;v^2(2k_nu)\;F^{abc}_2(u) \right], 
\label{eq9}
\end{equation}
\end{widetext}
where $v^2(2k_n\;u) =4(\gamma_1 + \gamma_2)+(8\gamma_3+4\gamma_4) 
(2k_n\;u)^2 + 4(\gamma_5+\gamma_6) (2k_n\;u)^4$, and 
all the other parameters are defined above.
The expressions of the three functions $F^{abc}_1(u)$, $F^{abc}_2(u)$ and $F^{abc}_3(u)$ are provided in Appendix B. 

To summarize, let us write the second-order energy correction in a compact form: 
\begin{eqnarray}
\nonumber 
\frac{\Delta E^{(2)}}{A}(\delta, \rho, \Lambda) &=& 
\sum_{i=n,p}\frac{\Delta E_{ii}+\Delta E^x_{ii}}{A}(\delta,\rho,\Lambda)+ \frac{\Delta E_{np}}{A} (\delta,\rho,\Lambda) \\
&=& \sum_{j=1}^{15} \chi_j (\delta, \rho) I_j(\delta,\rho,\Lambda). 
\label{eq10}
\end{eqnarray}

The last 5 terms in the above expression describe the $np$ contributions. For $j=11, 
\ldots ,15$ one has 
\begin{equation}
I_j(\delta,\rho,\Lambda)=\frac{1}{15} \int_0^{\frac{\Lambda}{2k_n}}du\;f_j(u)\; 
F^{abc}_{total}(u,\rho,\delta), 
\label{eq11}
\end{equation}
where

\vspace{0.3cm}
\begin{center}
\begin{tabular}{ccccc}
\hline
  $f_{11}(u)$ \;\;&\;\; $f_{12}(u)$ \;\;&\;\; $f_{13}(u)$ \;\;&\;\; $f_{14}(u)$ \;\;&\;\; $f_{15}(u)$ \\
\hline
 $u$ \;\;&\;\; $u^3$ \;\;&\;\; $u^3$ \;\;&\;\; $u^5$ \;\;&\;\; $u^5$ \\
\hline
\end{tabular}
\end{center},

\vspace{0.3cm}
and 
\begin{widetext}
\begin{equation}
F^{abc}_{total}(u,\rho,\delta)= \frac{1}{(bc)^3}\left[ \Theta(a-u) F^{abc}_1(u) + \Theta (u-a) 
\Theta (1-u) F^{abc}_3 (u) + \Theta (u-1) F^{abc}_2 (u) \right]. 
\label{eq12}
\end{equation}
\end{widetext}

The last 5 coefficients $\chi_j$ are equal to 
\begin{eqnarray}
\nonumber
\chi_{11}(\delta,\rho) &=& 32 \pi^3 C_{\Delta E} m^*_s k^7_n (t_{03}^2 + 
x_{03}^2 + t_{03} x_{03}),  \\
\nonumber
\chi_{12} (\delta,\rho) &=&  256 \pi^3 C_{\Delta E} m^*_s k^9_n (t_{03} t_{12} + 
x_{03} x_{12}),     \\
\chi_{13} (\delta,\rho) &=& 128 \pi^3 C_{\Delta E} m^*_s k^9_n (t_{03} x_{12}+ 
x_{03} t_{12}),       \\
\nonumber 
\chi_{14} (\delta,\rho) &=&   512 \pi^3 C_{\Delta E} m^*_s k^{11}_n t_{12} 
x_{12},      \\
\nonumber 
\chi_{15} (\delta,\rho) &=& 512 \pi^3 C_{\Delta E} m^*_s k^{11}_n (t_{12}^2 + 
x_{12}^2).  
\label{eq13}
\end{eqnarray} 

Starting from Eq. (20) the asymptotic behavior of the second-order energy contribution can be obtained by taking its asymptotic 
expansion. It can be shown that this leads to 
\begin{equation}
\frac{\Delta E^{(2)}}{A}(\delta, \rho, \Lambda \rightarrow \infty) = 
a^1_{\delta,\rho} \Lambda^5 + a^2_{\delta,\rho} \Lambda^3 + a^3_{\delta,\rho} \Lambda + a^4_{\delta,\rho} + O\left(\frac{k_F}{\Lambda}\right),
\label{eq14}
\end{equation}
where the coefficients $a^i$ depend on $\delta$ and $\rho$. 
We observe that the energy correction diverges as $\Lambda^5$ 
for large values of $\Lambda$ and that this divergence 
(expected from power counting arguments)
is much stronger than the linear divergence of the $t_0-t_3$ model, Eq. (1). 

Finally, the Skyrme EoS up to second order is written as 

\begin{widetext}
\begin{eqnarray}
\nonumber
\frac{E}{A}(\delta,\rho,\Lambda) &=& 
\frac{3\hbar^2}{10m}\left(\frac{3\pi^2}{2}\rho \right)^{\frac{2}{3}} G_{5/3} + 
\frac{1}{8} t_0 \rho [2(2+x_0)-(1+2x_0)G_2] 
+ \frac{1}{48} t_3 \rho^{\alpha+1} [2(2+x_3)-(1+2x_3)G_2] \\ &+& 
\frac{3}{40}\left(\frac{3\pi^2}{2}\right)^{\frac{5}{3}}
\rho^{\frac{5}{3}}\left[ \Theta_v G_{5/3} + \frac{1}{2} (\Theta_s -2 \Theta_v) 
G_{8/3} \right] + \sum_{j=1}^{15}\chi_j(\delta,\rho) I_j(\delta, \rho, \Lambda). 
\label{eq25}
\end{eqnarray}
\end{widetext}

\section{Results}

\subsection{Symmetric matter ($\delta=0$) and incompressibility modulus}

The EoS for symmetric matter is given by Eq. (25) 
by setting $\delta=0$. Let us also introduce the pressure and the incompressibility modulus 
calculated up to second order,  

\begin{widetext}
\begin{equation}
P(\rho,\Lambda)=\rho^2 \frac{d}{d\rho} \frac{E}{A}(\rho,\Lambda) = P^{(1)}(\rho) + 
\rho^2 \left[ \sum_{j=1}^{15} \left( \chi_{j}^{'} (\delta,\rho) I_j(\delta,\rho,\Lambda) + \chi_{j}(\delta,\rho)
I^{'}_j (\delta,\rho,\Lambda) \right) \right]_{\delta=0},
\label{eq16}
\end{equation}

\begin{eqnarray}
\nonumber
K_{\infty}(\rho, \Lambda) &=& 9\rho^2 \frac{d^2}{d\rho^2} \frac{E}{A}(\rho, \Lambda) \\ &=& K_{\infty}^{(1)}(\rho)   
+9\rho^2\left[\sum_{j=1}^{15}\chi_{j}^{''}(\delta,\rho)I_{j}(\delta,\rho,\Lambda)
+2\chi_{j}^{'}(\delta,\rho)I_j^{'}(\delta,\rho,\Lambda)
+\chi_{j}(\delta,\rho) I_j^{''}(\delta,\rho,\Lambda)\right]_{\delta=0},
\label{eq17}
\end{eqnarray}
\end{widetext}
where $P^{(1)}$ and $K^{(1)}$ denote the mean-field (first-order) pressure and the incompressibility modulus, respectively. 
The dependence on the cutoff appears in the second-order corrections. The mean-field expressions depend on 
the density $\rho$ \cite{sly5,meyer} as follows,   
\begin{equation}
P^{(1)}(\rho)=\rho \left[\frac{1}{5} \frac{\hbar^2}{m} \left(\frac{3\pi^2}{2}\right)^{2/3} \rho^{2/3} + 
\frac{3}{8} t_0 \rho +\frac{1}{16} \left(\frac{3\pi^2}{2}\right)^{2/3} \Theta_s \rho^{5/3}+ \frac{1}{16} 
t_3 (\alpha+1) \rho^{\alpha+1} \right]
\end{equation}
and
\begin{equation}
K_{\infty}^{(1)}(\rho)= -\frac{3}{5} \frac{\hbar^2}{m} \left(\frac{3\pi^2}{2}\right)^{2/3} \rho^{2/3} +
\frac{3}{8} \left(\frac{3\pi^2}{2}\right)^{2/3} \Theta_s \rho^{5/3} +
\frac{9}{16}\alpha (\alpha+1) t_3 \rho^{\alpha+1}.
\end{equation} 
The $'$ notation in Eqs. (26) and (27) denotes the derivative with respect to the density $\rho$.\\
In the upper panel of Fig. 2 we plot the second-order EoS obtained for different values of the cutoff $\Lambda$ (see legend), from 0.5 up to 2 fm$^{-1}$. The different equations of state are calculated by using the SLy5 Skyrme parameters and are compared with the reference mean-field SLy5 EoS (solid line in (a)). In (b) the second-order correction is plotted for the same values of the cutoff. 
We observe that for a cutoff value equal to 2 fm$^{-1}$ the correction to the energy at the saturation point of nuclear matter, $\rho=$ 0.16 fm$^{-3}$, is very large and amounts to - 80 MeV.  
By using the same values of the cutoff $\Lambda$ and the SLy5 parameters, the second-order pressure and the second-order incompressibility modulus are displayed in Fig. 3. 
One may observe how strongly the ultraviolet divergence affects the pressure and the incompressibility modulus for large values of the cutoff $\Lambda$. The incompressibility is strongly enhanced by the second-order correction and is equal to $\sim$ 625 MeV at the saturation point of matter for 
$\Lambda=2$ fm$^{-1}$. 

To have a reasonable second-order EoS, we have adjusted the nine parameters of the Skyrme interaction entering in the expression of the EoS to reproduce the reference SLy5 mean-field EoS. We have chosen 15 equidistant reference points ($N$) for densities ranging from 0.02 fm${}^{-3}$ to 0.30 fm${}^{-3}$. All the parameters are kept free in the adjustment procedure.
The minimization has been performed using the following definition for the $\chi^2$,
\begin{equation}
\chi^2=\frac{1}{N-1}\sum_{i=1}^{N} \frac{(E_i-E_{i,ref})^2}{\Delta E_i^2}.
\end{equation}
The errors or {\it adopted} standard deviations, $\Delta E_i$, in Eq. (30) are chosen equal to 1\% of the reference SLy5 mean-field energies $E_{i,ref}$. This choice is arbitrary since we are fitting a theoretical EoS where a standard deviation for this quantity has not been estimated. However, the magnitude of the $\chi^2$ defined in Eq. (30) has a clear and reasonable meaning: if it is smaller or equal to one, the reference EoS is reproduced within one standard deviation, i.e., within a 1\% average error by our second-order EoS. The corresponding curves obtained with the adjusted parameters are shown in Fig. 4 for different values of $\Lambda$. The quantities which are displayed in this figure are the differences between the refitted EoS and the reference SLy5 mean-field EoS for different values of the cutoff $\Lambda$. We observe that the deviations are extremely small except at very low densities where they are anyway not larger than 0.06 MeV. In the inset of the figure the refitted EoS are plotted and compared with the SLy5 mean-field EoS (solid line). Due to the scale, the curves in the inset  are practically indistinguishable. The obtained sets of parameters and the $\chi^2$ values are shown in Table I for each value of the cutoff $\Lambda$. The $\chi^2$ values are always extremely small indicating that, on average, the fitted points are deviating much less than 1 \% (according to the adopted expression for $\chi^2$, Eq. (30)) with respect to the reference EoS. 
 
We have noticed that, for the four refitted interactions, the saturation density $\rho_0$ and the incompressibility modulus are equal in all cases to 0.16 fm$^{-3}$ and 229.9 MeV, respectively. 

The pressure and the incompressibility, 
Eqs.~(\ref{eq16}) and (\ref{eq17}), evaluated by using the parameters listed in Table I, are plotted in the two panels of Fig. 5. Again, what is plotted is the deviation with respect to the SLy5 mean-field reference values. In the two insets, the absolute values are displayed together with the SLy5 mean-field curves (solid lines). We stress that the pressure and the incompressibility do not enter in the fits. In spite of this, small deviations from the SLy5 reference curves are observed, only at large densities. 

\subsection{Pure neutron matter ($\delta=1$)}

By setting $\delta=1$ in Eq. (25) the mean-field plus second-order EoS is obtained for pure neutron matter. 
The ultraviolet divergence with respect to the cutoff is visible in Fig. 6 where the EoS (a) 
and the second-order correction (b) are displayed for different values of the cutoff $\Lambda$. 
In the upper panel the reference SLy5 mean-field EoS is also plotted (solid line). 
One 
notices a special and unexpected behavior: starting 
from the cutoff value $\Lambda=1.5$ fm$^{-1}$ the corrected 
EoS has an equilibrium point and, for $\Lambda=2$ 
fm$^{-1}$, the total energy is 
negative. 
The appearance of an equilibrium point for the second-order EoS of pure neutron matter shows how the ultraviolet divergence is also responsible for generating artificial (and unphysical) strong correlations in the system. This anomaly can be cured by the adjustment of the parameters. 
We have performed also in this case the adjustment of the nine parameters of the Skyrme interaction with the same definition of $\chi^2$ as above, Eq. (30). The fitted points are the same as in the previous case. In Fig. (7) the deviations with respect to the SLy5 mean-field EoS are shown. 
Again, the deviations are extremely small and are larger at very low densities. 
In the inset the absolute curves are plotted. The obtained parameters and the $\chi^2$ values per point are listed in Table II. The $\chi^2$ values are extremely small also in this case and not larger than 10$^{-6}$.

\subsection{An illustration of asymmetric matter ($\delta=0.5$)}

We have chosen the asymmetry value $\delta=0.5$ to illustrate a case of asymmetric matter. 
The corrected EoS and the second-order correction are presented in the upper and 
lower panels of Fig. 8, respectively. The results of the fit (same definition of $\chi^2$ and same number of fitted points as for the other two cases) are shown in Fig. 9 whereas the 
sets of parameters are listed in Table III. 
The quality of the fit is very good also in this case as indicated by the low values of $\chi^2$. These values have increased with respect to the   
two previous adjustments but they still remain much lower than 1.

\subsection{Global fit for the three values $\delta$ = 0, 0.5 and 1}

Finally, a unique and global fit has been done to readjust the three 
mean-field plus second-order EoS for symmetric, asymmetric and 
pure neutron matter
to reproduce the corresponding SLy5 mean-field curves.  
The obtained sets of parameters are presented in Table IV. 
In Fig. 10 the three refitted EoS are plotted 
and in Fig. 11 the deviations from the Sly5 mean-field curves are shown. The resulting pressure and 
incompressibility modulus for symmetric matter are shown in Fig. 12. Their deviations with respect to the SLy5 mean-field values are presented in Fig. 13. 

Globally, as one can see from the $\chi^2$ values, the quality of the fit is deteriorated with respect to that found for each separate case. 
However, the fit is still of good quality.
The $\chi^2$ in this global case is composed by the three contributions as calculated in the previous subsections and the final value is divided by three in order to make our different results comparable to one another. Specifically, the $\chi^2$ values are still less than 1 up to $\Lambda=$ 1 fm$^{-1}$. Values between 1 and 2 (to be judged by considering the adopted choice of the errors in the expression of $\chi^2$) are found for larger values of the cutoff meaning that the fit is still good. 
 
We have considered in our global fit only three values of the asymmetry parameter $\delta$, describing symmetric, neutron and one case of asymmetric matter, $\delta=0.5$. To judge the quality of the refitted parameters also for other values of $\delta$ (other values of the asymmetry) we show in Figs. 14 and 15 a test performed with the parameters which have been obtained from the global fit. Two values of the density (lower and larger than the saturation density) have been chosen and the deviations between the refitted EoS and the reference SLy5 mean-field EoS (for different values of the cutoff) are plotted as a function of $\delta$ for $\rho=0.1$ (Fig. 14) and 0.2 (Fig. 15) fm$^{-3}$, respectively. We observe that the deviations are always reasonably small for all the values of $\delta$ and (which is the most important result of this test) that they do not increase strongly for the values of $\delta$ which have not been used in the fit. The maximum deviations are not larger than 0.4 MeV.  

The values of the saturation density and of the incompressibility modulus for symmetric matter resulting from the global fit are displayed in  Table V for the four values of the cutoff $\Lambda$.

Finally, for the case of the global fits that constitute our more demanding test to the second-order EoS, we have estimated the standard deviation of the fitted parameters \cite{bevington}. This analysis allows one to asses how well the used reference data toghether with the adopted errors constraint the parameters of our model. In particular, the standard deviation associated to such parameters are displayed in Table IV.

\section{Conclusions}

We have analyzed in this work the nature of the ultraviolet divergence generated by the zero range of the Skyrme interaction in the second-order EoS of nuclear matter. The same issue has been previously addressed \cite{mog} but in a simple $t_0-t_3$ model and by considering only symmetric nuclear matter. A cutoff regularization has been proposed in Ref. \cite{mog}
to treat the ultraviolet divergence (which was linear with the momentum cutoff $\Lambda$). 
In this work the velocity-dependent terms of the Skyrme interaction 
have also been included and both symmetric and asymmetric matter 
have been considered, including the extreme case of pure neutron matter. 
The expressions of the second-order correction to the EoS are 
derived analytically and a strong divergence ($\sim$
$\Lambda^5$) is found in the asymptotic expression of the corrective terms. A cutoff regularization procedure is adopted first for the single cases of symmetric, neutron and asymmetric ($\delta=0.5$) matter. The resulting fits are of extremely good quality. A global fit is finally performed simultaneously for the three EoS. The results are still satisfactory. 

Two interesting conclusions may be drawn: i) Even if the divergence is 
much stronger than in the simple $t_0-t_3$ case, the fit of the parameters is still 
possible; 
ii) The three EoS may be adjusted simultaneously and the problem 
of the appearance of an artificial equilibrium point for neutron matter can be always cured by the adjustment of the parameters. 

The adjusted interactions display reasonable properties for nuclear matter. 
 This opens new perspectives for future applications of this kind of interactions in 
beyond mean-field models to treat finite nuclei. 
It is worth reminding that, so far, conventional phenomenological 
interactions (adjusted at the mean-field level) have been 
employed for nuclei in different beyond-mean-field calculations 
 \cite{colo,gambacurta,pillet}. 
   
A drawback of the cutoff regularization procedure is the fact that for each momentum cutoff a different parametrization is generated. A unique set of parameters could be provided by applying the dimensional renormalization (mentioned in Sec. I). Work to apply the dimensional renormalization to the second-order EoS of nuclear matter is in progress.

\vspace{1cm}

\noindent{\bf Acknowledgments}
The authors thank Nguyen van Giai for fruitful discussions. 
This work is  supported in part by the
Italian Research Project "Many-body theory of
nuclear systems and implications on the
physics of neutron stars" (PRIN 2008).

\newpage

\begin{figure}
\includegraphics[width=9cm]{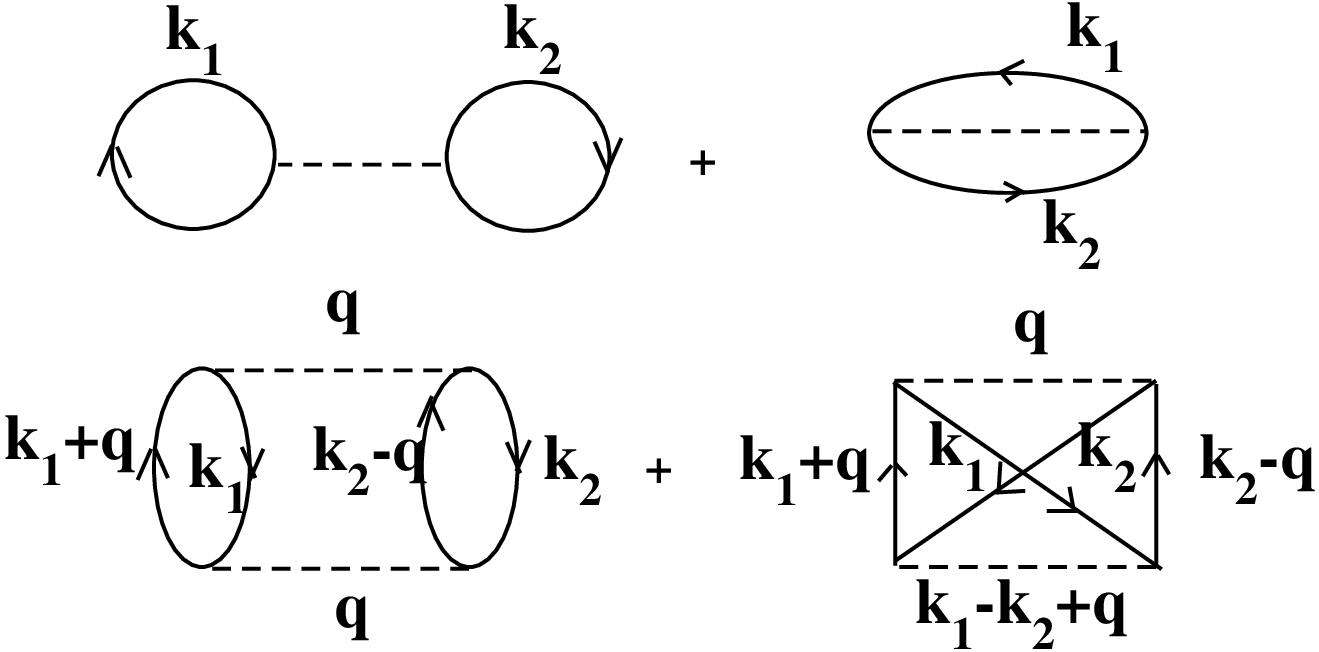}
\caption{Direct and exchange first-order (upper line) and second-order (lower line) contributions to the total energy.}
\label{fig1}
\end{figure}

\vspace{1cm}

\begin{figure}
\includegraphics[width=12cm]{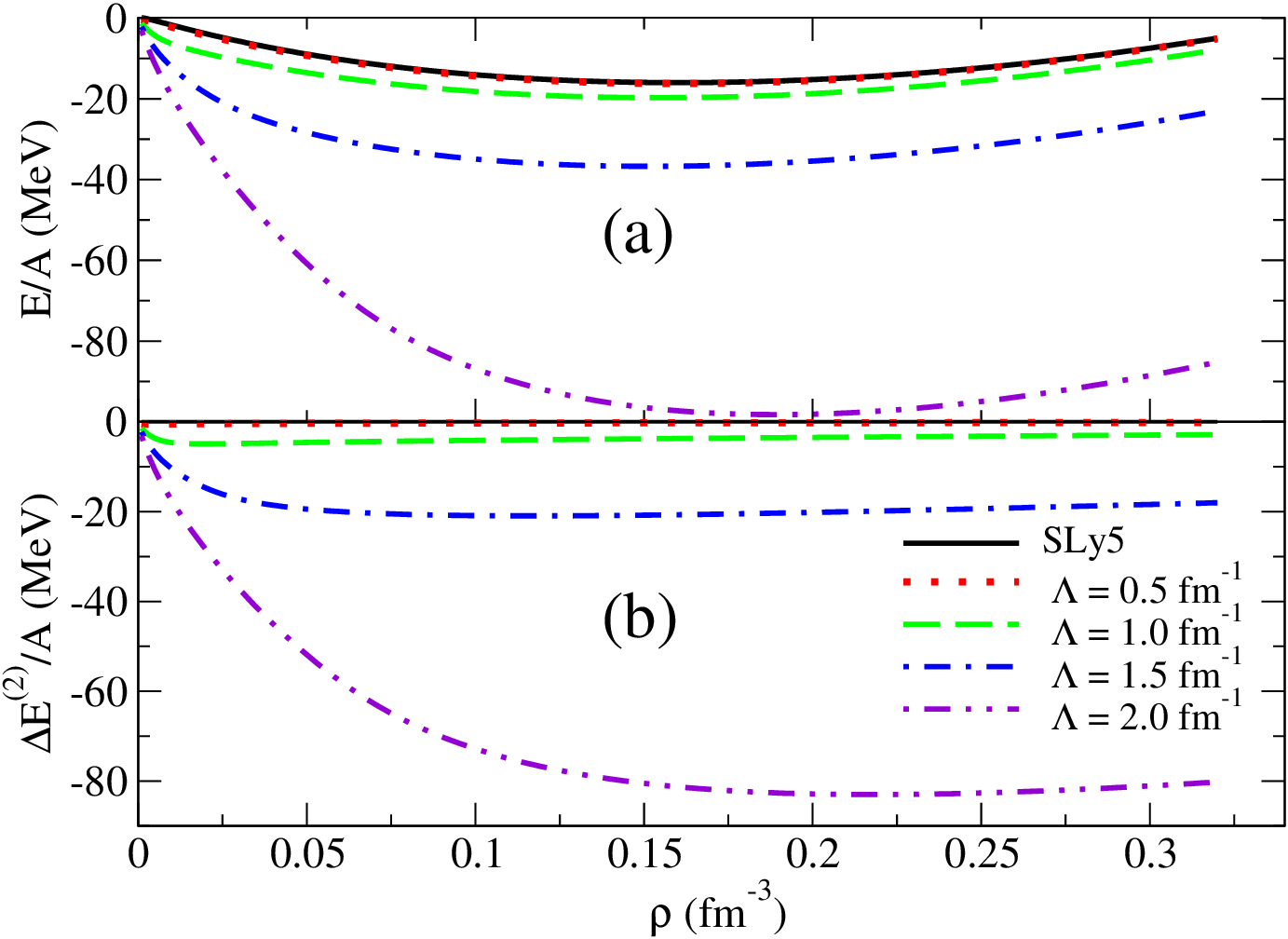}
\caption{(Color online) (a) Second-order EoS for different values of the cutoff $\Lambda$ and (b)  
second-order correction  for symmetric nuclear matter calculated with the SLy5 parameters. The SLy5 mean-field EoS is also plotted in (a) (solid line).}
\label{fig2}
\end{figure}

\vspace{1cm}

\begin{figure}
\includegraphics[width=12cm]{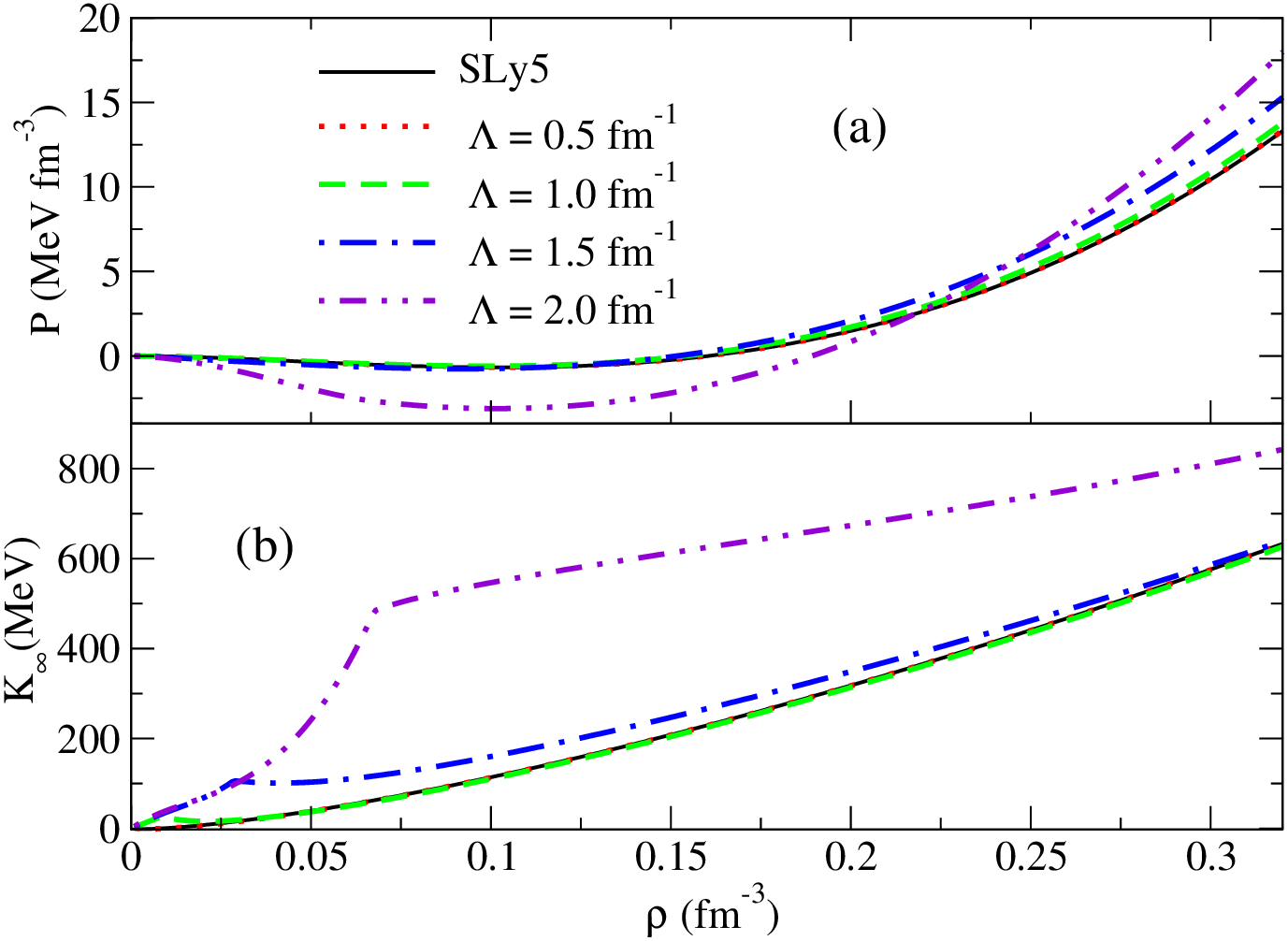}
\caption{(Color online) Second-order pressure (a) and incompressibility modulus (b) calculated with the SLy5 parameters for different values of the cutoff. The mean-field SLy5 curves are also plotted in both panels (solid lines).}
\label{fig3}
\end{figure}

\vspace{1cm}

\begin{figure}
\includegraphics[width=12cm]{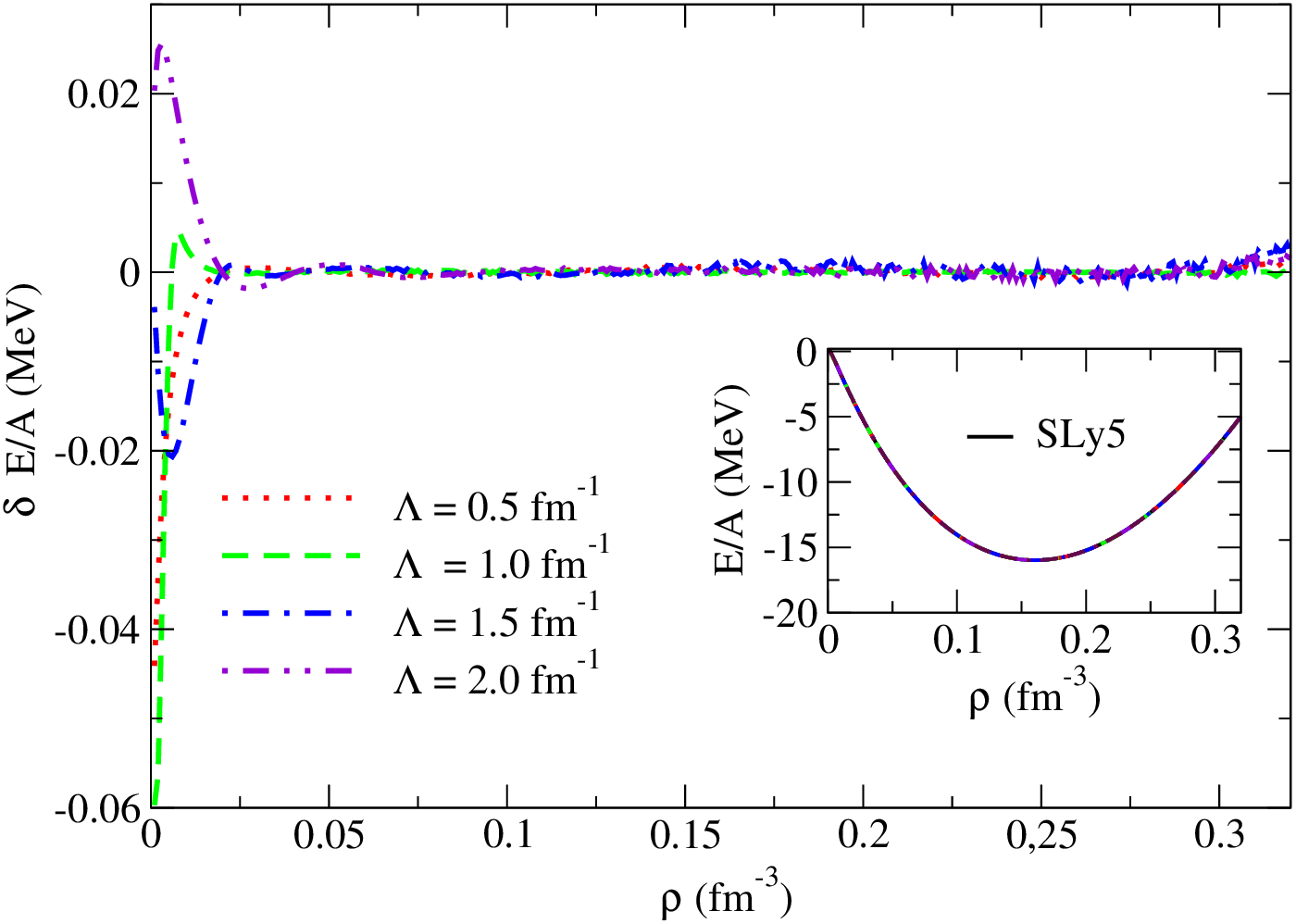}
\caption{(Color online) Deviations between the refitted second-order EoS (for different values of the cutoff) and the SLy5 mean-field curve for symmetric nuclear matter. In the inset the absolute values are plotted and compared with the SLy5 mean-field EoS (solid line).}
\label{fig4}
\end{figure}

\vspace{1cm}

\begin{figure}
\includegraphics[width=12cm]{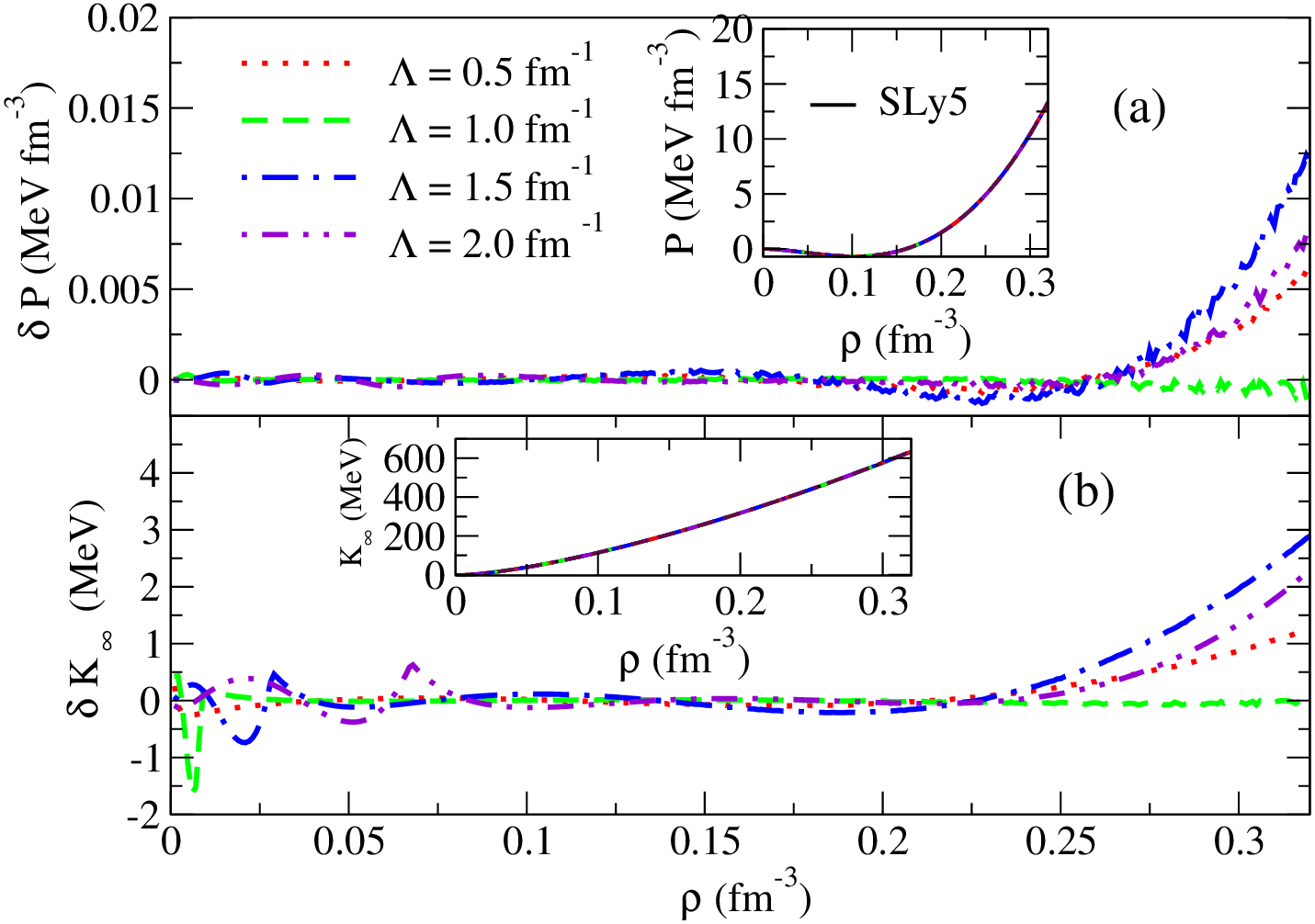}
\caption{(Color online) Deviations of the pressure (a) and of the incompressibility (b) (calculated with the refitted parameters for symmetric 
nuclear matter) with respect to the mean-field SLy5 values. In the insets the absolute values are displayed and compared with the SLy5 mean-field curves (solid lines).}
\label{fig5}
\end{figure}

\vspace{1cm}

\begin{figure}
\includegraphics[width=12cm]{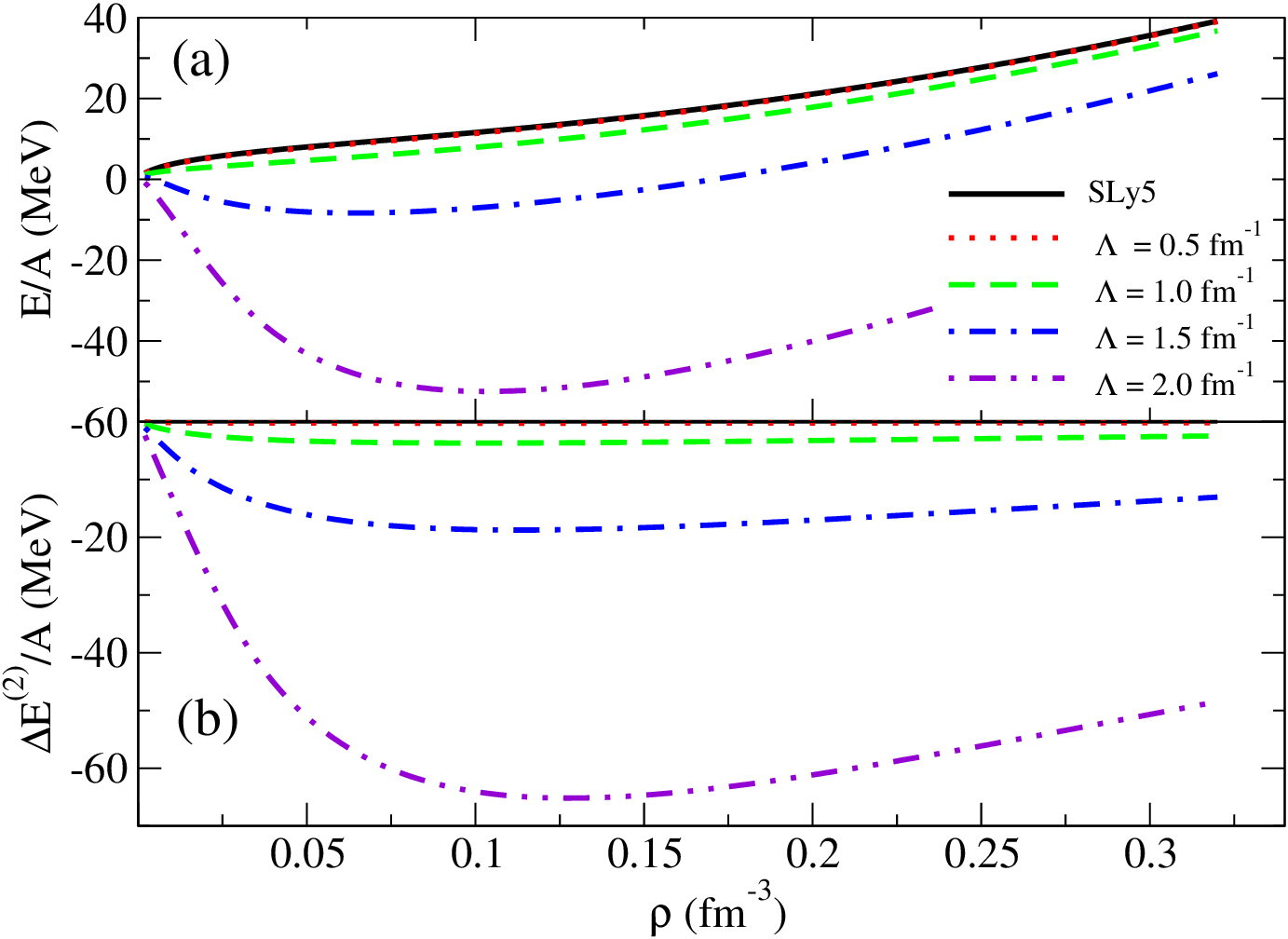}
\caption{(Color online) As in Fig. 2 but for pure neutron matter.}
\label{fig6}
\end{figure}

\vspace{1cm}

\begin{figure}
\includegraphics[width=12cm]{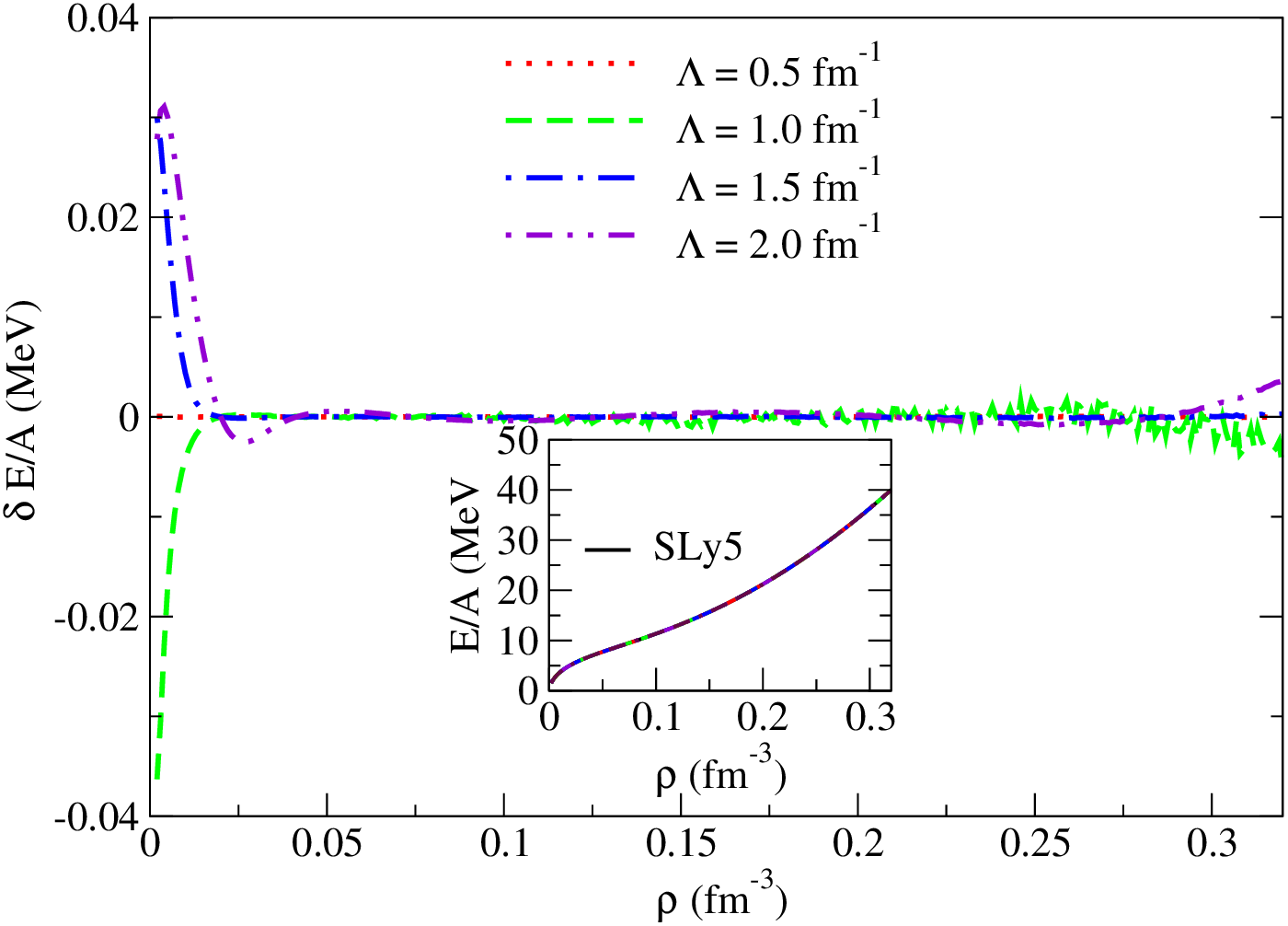}
\caption{(Color online) Deviations of the refitted EoS for pure neutron matter with respect to the SLy5 mean-field EoS. In the inset the absolute curves are displayed with the SLy5 mean-field EoS (solid line).}
\label{fig7}
\end{figure}

\vspace{1cm}

\begin{figure}
\includegraphics[width=12cm]{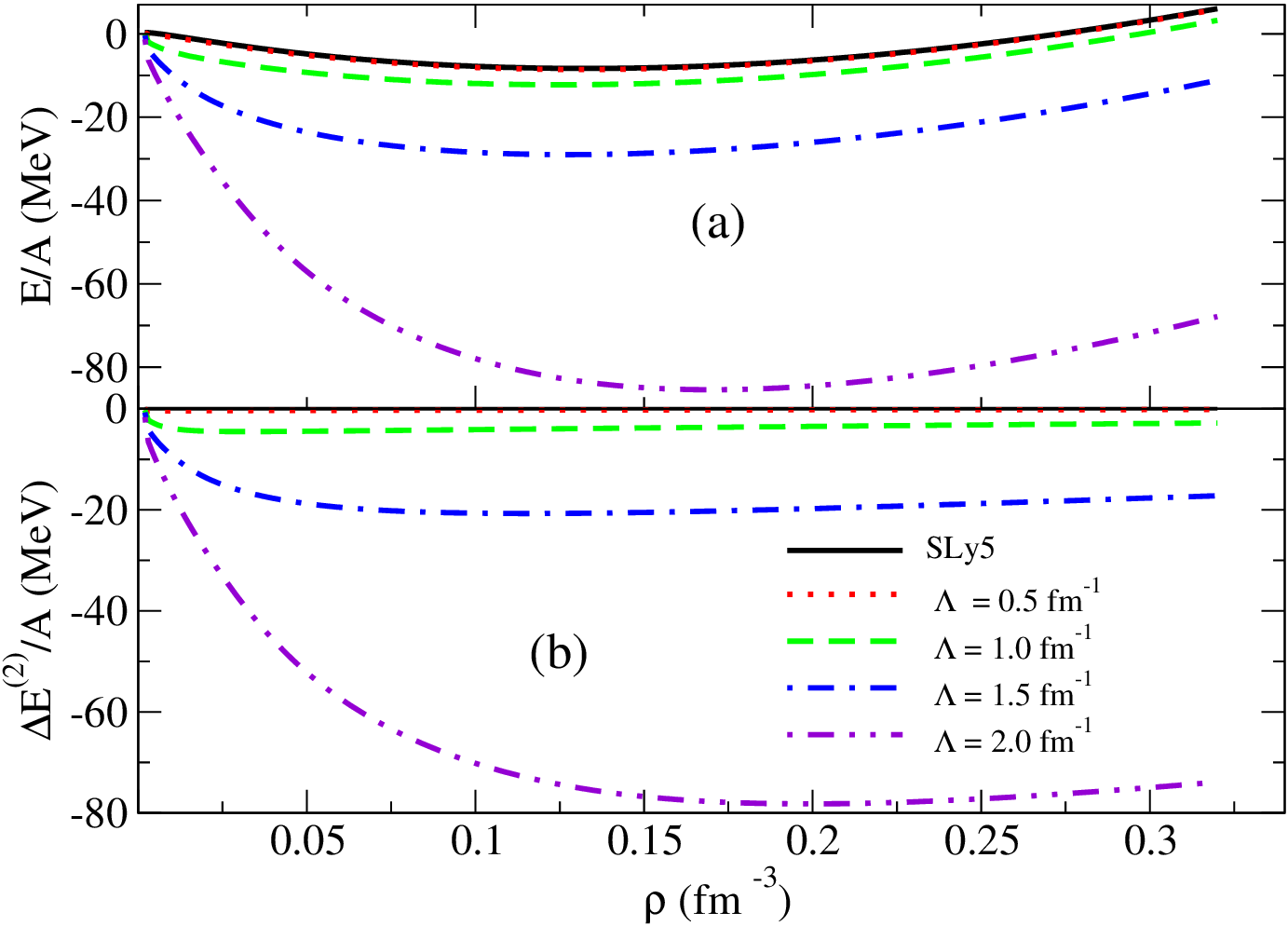}
\caption{(Color online) As in Fig. 2 but for asymmetric nuclear matter in the case $\delta=0.5$.}
\label{fig8}
\end{figure}

\vspace{1cm}

\begin{figure}
\includegraphics[width=12cm]{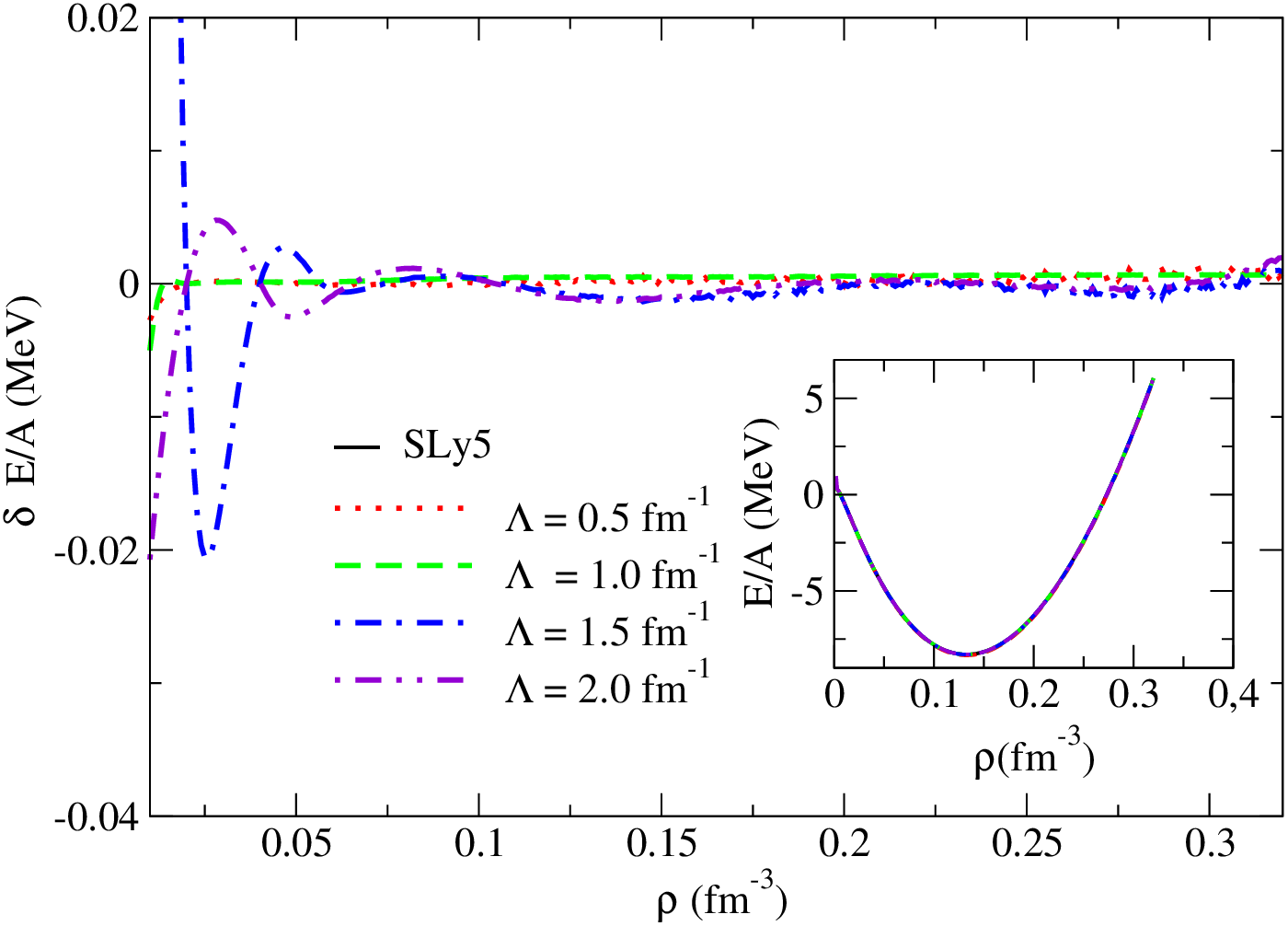}
\caption{(Color online) As in Fig. 7 but for asymmetric matter ($\delta=0.5$).}
\label{fig9}
\end{figure}

\vspace{1cm}

\begin{figure}
\includegraphics[width=12cm]{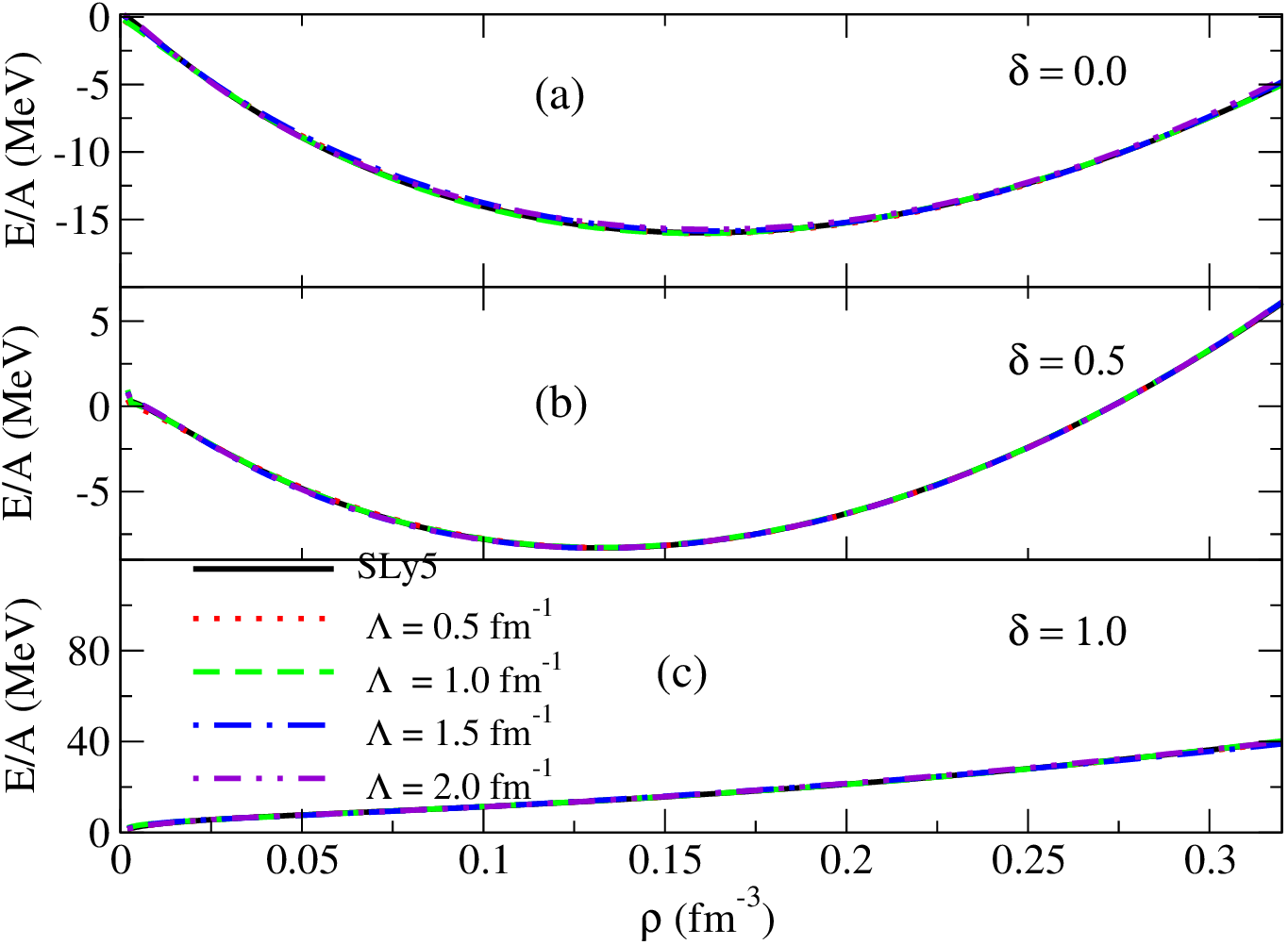}
\caption{(Color online) Refitted EoS (global fit) for symmetric (a), asymmetric (b) 
and pure neutron (c) matter. The reference SLy5 mean-field curves are also plotted in the 3 panels (solid lines).}
\label{fig10}
\end{figure}

\vspace{1cm}

\begin{figure}
\includegraphics[width=12cm]{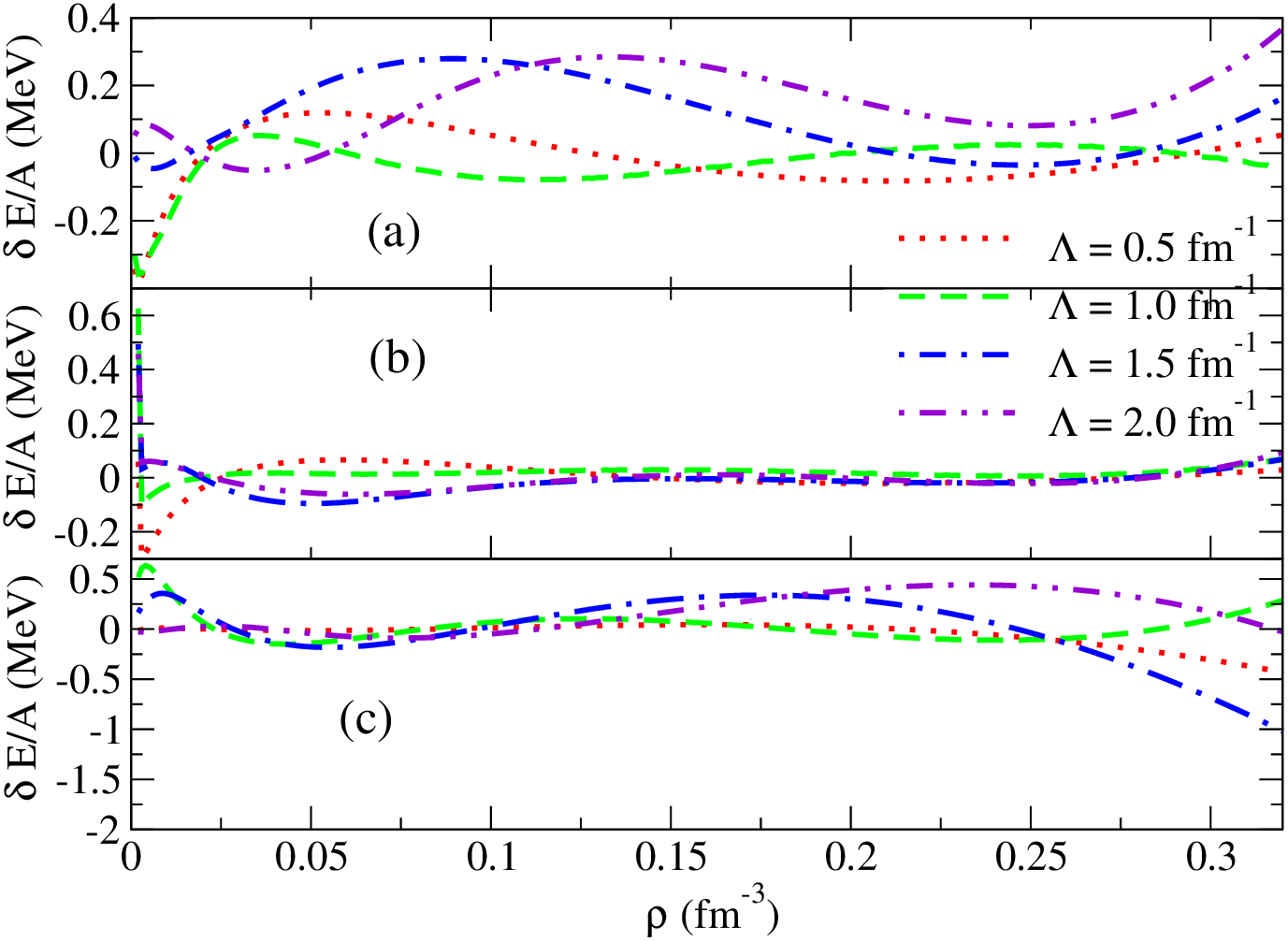}
\caption{(Color online) Deviations of the refitted EoS (global fit) for symmetric (a), asymmetric (b) 
and pure neutron (c) matter with respect to the mean-field SLy5 values.}
\label{fig10}
\end{figure}

\vspace{1cm}
\begin{figure}
\includegraphics[width=12cm]{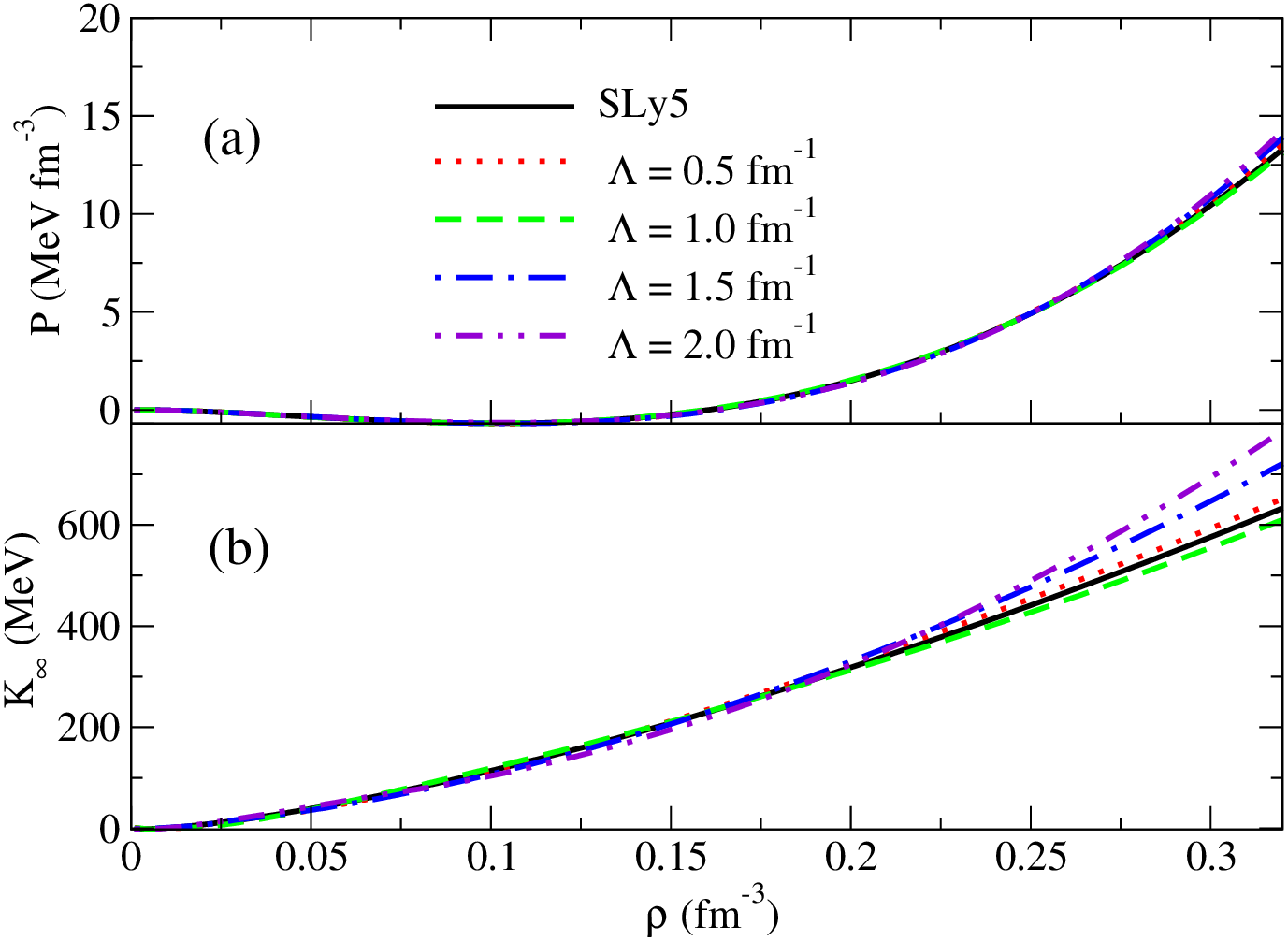}
\caption{(Color online) Pressure (a) and incompressibility (b) evaluated with the parameters obtained with the global 
fit.}
\label{fig11}
\end{figure}

\vspace{1cm}
\begin{figure}
\includegraphics[width=12cm]{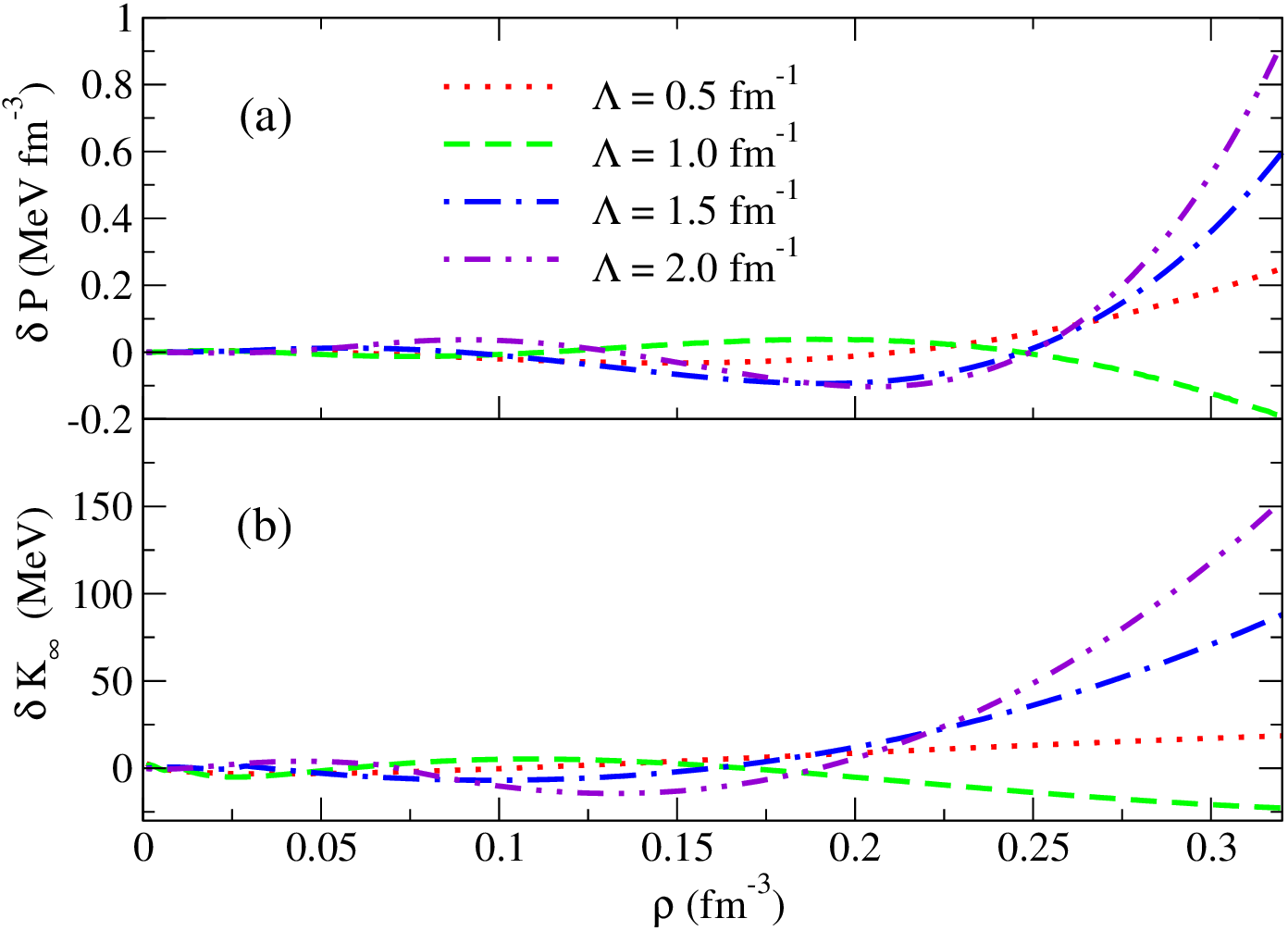}
\caption{(Color online) Deviations of the pressure (a) and of the incompressibility (b) (evaluated with the parameters obtained with the global 
fit) with respect to the mean-field SLy5 curves.}
\label{fig11}
\end{figure}

\vspace{1cm}
\begin{figure}
\includegraphics[width=12cm]{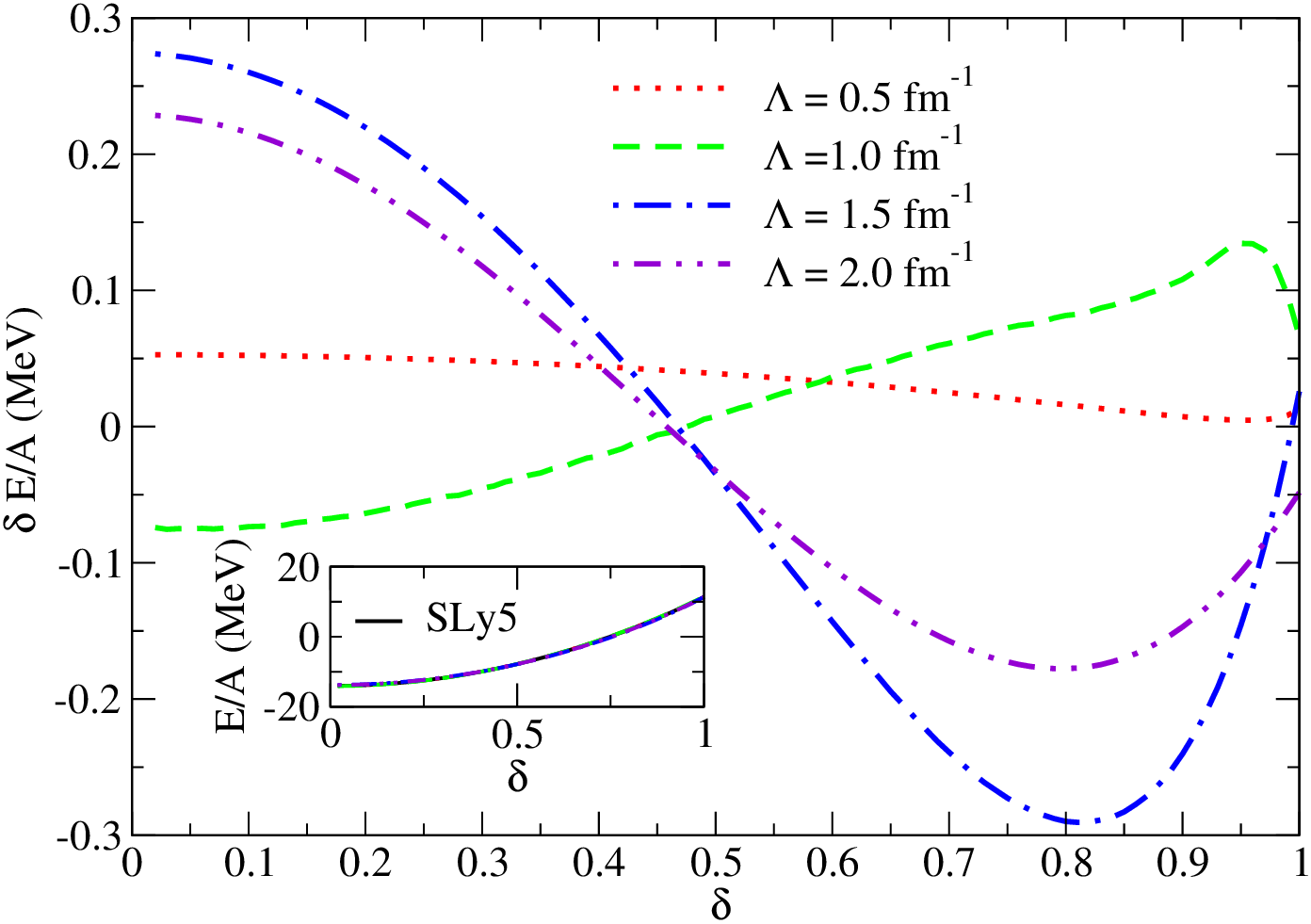}
\caption{(Color online) Deviation between the refitted EoS and the reference SLy5 mean-field EoS as a function of $\delta$ for  $\rho = $ 0.1 fm$^{-3}$. In the inset the absolute values are plotted.}
\label{fig11}
\end{figure}

\vspace{1cm}
\begin{figure}
\includegraphics[width=12cm]{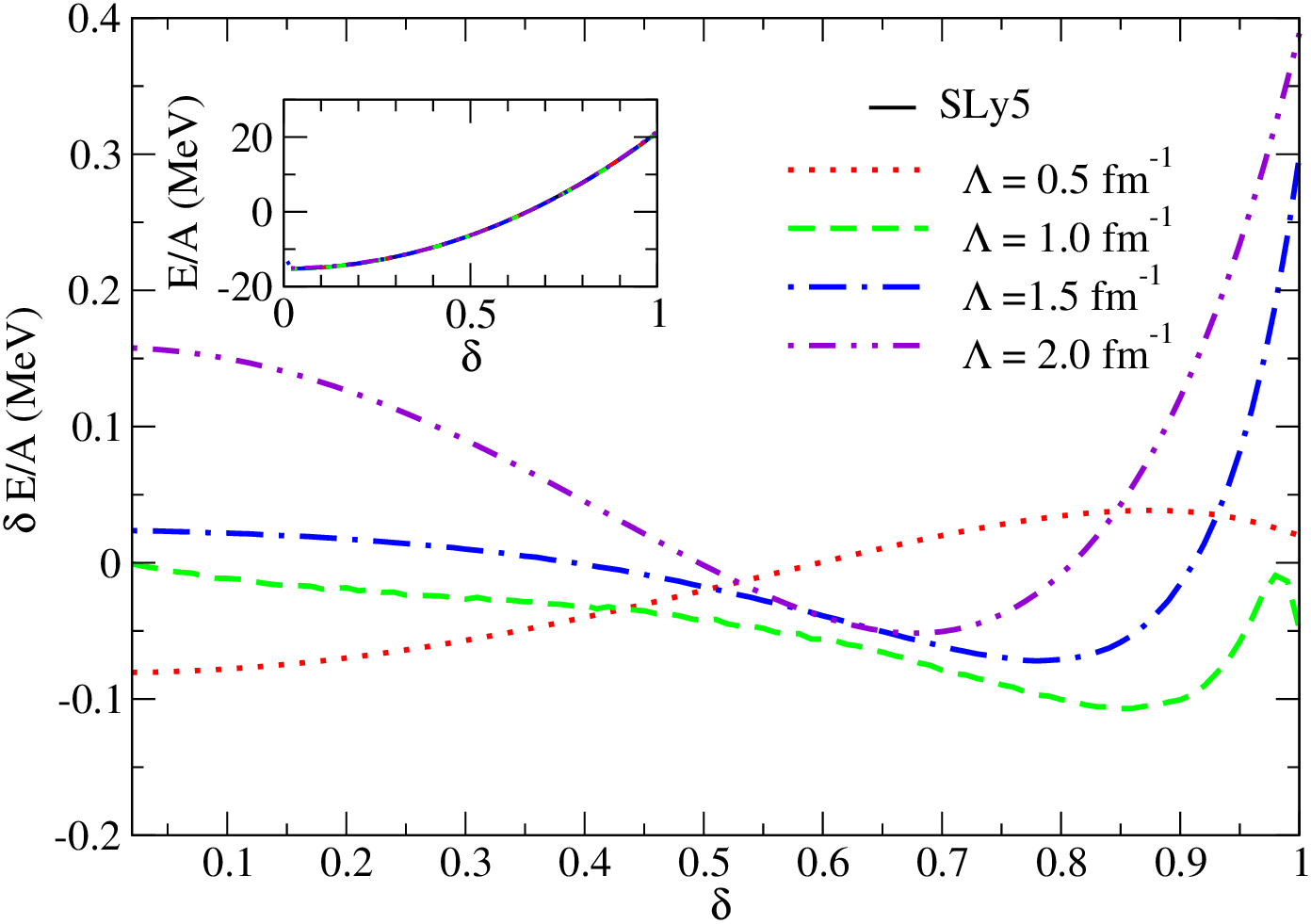}
\caption{(Color online) Same as in Fig. 15 but for $\rho = $ 0.2 fm$^{-3}$. }
\label{fig11}
\end{figure}

\vspace{1cm}

\begin{table}[!h]
\centering
\caption{Parameter sets obtained in the fit of the EoS of symmetric matter for different values of the cutoff $\Lambda$ compared with the original set SLy5. In the last column the $\chi^2$ values are shown. }\vspace{0.5cm}\label{symmetric_parameters}
\begin{tabular}{  c c c c c c c c c c c }
    \hline
    \hline
  &    $\quad t_0$ & $t_1$ & $t_2$ & $t_3$ & $x_0$ & $x_1$ & $x_2$ & $x_3$ & $\alpha$ &  \\
 & (MeV fm$^3$) & (MeV fm$^5$) & (MeV fm$^5$) & (MeV fm$^{3+3\alpha}$) & & & & & &  \\
SLy5 &-2484.88    &    483.13   &   -549.40   & 13763.0&0.778     &    -0.328    &  -1.0 &1.267   &   0.16667  &\\   
\hline
  $\Lambda$(fm$^{-1})$ & &  &  &  &  &  &  &  & & $\chi^2$   \\
 0.5 & -1817.280  &  646.948 & 4373.135 & 10101.307   &   -0.0002 & -3.464  & -1.314  & 6.233   &   0.246 &   6.2e-06 \\
1.0 & -1132.001 &  807.361 & -323.413  &  7555.400  &  0.733  &  1.201 & 0.644  &  5.012  &  0.457 &  4.9e-07 \\
1.5 & -608.125 & 71.647  & 241.517 & -3920.616  &  1.565  & -2.376 & 1.655  & 6.111   &  0.834 &  7.2e-06  \\
2.0 & -331.658 & 660.677 & -695.979 & -90.060 & 3.000 & -0.803  & -1.120 & 164.031   &  0.754 & 3.3e-06   \\
   \hline\hline
  \end{tabular}
  \end{table}

\vspace{1cm}

\begin{table}[!h]
\centering
\caption{Parameter sets obtained in the fit of the EoS of neutron matter for different values of the cutoff $\Lambda$ compared with the original set SLy5. In the last column the $\chi^2$ values are shown.  }\vspace{0.5cm}\label{symmetric_parameters}
\begin{tabular}{  c c c c c c c c c c c  }
    \hline
    \hline
   & $\quad t_0$ & $t_1$ & $t_2$ & $t_3$ & $x_0$ & $x_1$ & $x_2$ & $x_3$ & $\alpha$ &  \\
 & (MeV fm$^3$) & (MeV fm$^5$) & (MeV fm$^5$) & (MeV fm$^{3+3\alpha}$) & & & & & &  \\
SLy5 &-2484.88	&483.13     &  -549.40	&   13736.0    &   0.778	&       -0.328	&     -1.0	&   1.267&	0.16667 & \\
\hline
$\Lambda$(fm$^{-1})$ & & & & & & & & & & $\chi^2$  \\
0.5 & -535.222   &  403.303  & 1660.746  &  42905.115   & 0.094   & -0.970 & -1.031   & 1.094  & 0.144 & 9.1e-09   \\
1.0 & -1941.276    & 92.989   & 393.422     & -137583.116   & 0.609   & -0.502  & -1.010  & 1.057  & 0.613 & 1.5e-06  \\
1.5 & -18033.283   &  319.198  & -186.907  & 110184.232  &  1.846   & -1.113 & -0.929 & 1.893 &  0.006 & 4.7e-08  \\
2.0 & -218.464   &   598.755  & -538.604  & 496.206  &  0.015   & -0.885 & -0.745 &  14.793 &  0.205 & 6.6e-06  \\
   \hline\hline
  \end{tabular}
  \end{table}

\vspace{1cm}

\begin{table}[!h]
\centering
\caption{Parameter sets obtained in the fit of the EoS of asymmetric matter ($\delta=0.5$) matter for different values of the cutoff $\Lambda$ compared with the original set SLy5. In the last column the $\chi^2$ values are shown. }\vspace{0.5cm}\label{symmetric_parameters}
\begin{tabular}{  c c c c c c c c c c c}
    \hline
    \hline
  &   $\quad t_0$ & $t_1$ & $t_2$ & $t_3$ & $x_0$ & $x_1$ & $x_2$ & $x_3$ & $\alpha$ &  \\
 & (MeV fm$^3$) & (MeV fm$^5$) & (MeV fm$^5$) & (MeV fm$^{3+3\alpha}$) & & & & & &  \\
SLy5 & -2484.88	&483.13    &   -549.40	& 13763.0  &    0.778	&    -0.328	&  -1.0	       &   1.267& 0.16667 & \\
\hline
$\Lambda$(fm$^{-1}$) & & & & & & & & & & $\chi^2$ \\
0.5 & -2691.295   &  2227.930 &  -275.173 & 19875.288 &  1.109   & -1.510  & 4.268 & 2.790 &   0.116 & 2.5e-04    \\
1.0 & -4139.692   & 771.130 &  1079.952 &  20372.212 &  -1.159   & 2.114  & -1.047 &  -1.790 &  0.027 & 7.0e-04 \\
1.5 & -1005.707 &  651.553 & -297.441 & 202.122  & 1.357  &  0.708  & -1.306 &  2.657 &  -0.434 & 3.5e-04 \\  
2.0 & -2795.987 & 699.587 & -563.067 & 11780.236 & 5.252 &  -0.515 & -0.939  & 7.119 & -0.007 & 2.1e-04 \\
   \hline\hline
  \end{tabular}
  \end{table}

\vspace{1cm}

\begin{table}[!h]
\centering
\caption{Parameter sets obtained in the fit of the EoS of symmetric, asymmetric and pure neutron matter for different values of the cutoff $\Lambda$ compared with the original set SLy5. The standard deviation, $\sigma$, estimated for the different parameters is also given. In the last column the $\chi^2$ values  are shown.}\vspace{0.5cm}\label{symmetric_parameters}
\begin{tabular}{  c c c c c c c c c c c }
    \hline 
    \hline
   &  $\quad t_0$ & $t_1$ & $t_2$ & $t_3$ & $x_0$ & $x_1$ & $x_2$ & $x_3$ & $\alpha$ &  \\
   &  $\quad \sigma_{t_0}$ & $\sigma_{t_1}$ & $\sigma_{t_2}$ & $\sigma_{t_3}$ & $\sigma_{x_0}$ & $\sigma_{x_1}$ & $\sigma_{x_2}$ & $\sigma_{x_3}$ & $\sigma_{\alpha}$ &  \\
 & (MeV fm$^3$) & (MeV fm$^5$) & (MeV fm$^5$) & (MeV fm$^{3+3\alpha}$) & & & & & &  \\
SLy5 & -2484.88	&483.13     &  -549.40	&   13736.0    &   0.778	&       -0.328	&     -1.0	&   1.267&	0.16667  &\\
\hline
$\Lambda$(fm$^{-1})$ & & & & & & & & &  &$\chi^2$ \\ 
0.5 & -2022.142 &  290.312 &  1499.483 &  12334.459 &  0.481 & -5.390 & -1.304 &  0.880 &  0.259 & 0.411 \\
    &  $ 0.49$ & $ 0.212$& $ 1.75$ & $ 4.5$ & $ 0.001173$ & $ 0.00657$ & $ 0.00020$ &  $ 0.001632$ & $ 0.000280$ &  \\
1.0 & -627.078 &  83.786 & -971.384 &  186.775  &  3.428 &  -1.252 & -1.620 & 200.360 & 0.338 & 0.540 \\
    &  $ 1.668$ & $ 0.2740$ &  $ 0.782$ & $ 0.078$ &  $ 0.00260$ & $ 0.01927$ & $ 0.00026$ & $ 0.082$ & $ 0.000314$ &  \\
1.5 & -743.227 &  112.246 & -42.816 &  5269.849 &  1.013 &  3.478 & -2.114 & 0.189 &  0.814 & 1.733 \\
    &  $ 0.306$ & $ 0.685$ &  $ 0.2972$ & $ 5.4$ &  $ 0.01415$ & $ 0.01309$ & $ 0.00519$ & $ 0.045037$ & $ 0.000784$ &  \\
2.0 & -718.397 &  573.884 & -497.766 &  6179.243 &  0.391 & -0.393 & -0.574 & 0.785  & 1.051 & 1.313   \\
    &  $ 0.343$ & $ 0.251$ &  $ 0.261$ & $ 8.33$ &  $ 0.005876$ & $ 0.001850$ & $ 0.000597$ & $ 0.017475$ & $ 0.00104$ &  \\
   \hline\hline
  \end{tabular}
  \end{table}

\vspace{1cm}

\begin{table}[!h]
\centering
\caption{ Saturation density and incompressibility modulus resulting from the global fit for symmetric nuclear matter.}\vspace{0.5cm}\label{saturation}
\begin{tabular}{ c c c }
    \hline 
$\Lambda$ (fm$^{-1}$) & $\rho_0$ (fm$^{-3}$) & $K_{\infty}$ (MeV) \\
\hline
0.5 & 0.16 & 236.36 \\
1.0 & 0.16 & 230.52 \\
1.5 & 0.16 & 236.28 \\
2.0 & 0.16 & 222.76 \\ 
\hline
  \end{tabular}
  \end{table}

\newpage

\section{APPENDIX A}
The expressions of the ten functions $F^j_1 (u)$ and $F^j_2 (u)$ (with $j$ running from 1 to 5) appearing in Eq. (16) read,  
\begin{eqnarray}
\nonumber
F_{1}^{1}(u)&=&\left(4+\frac{15}{2}u-5u^3+\frac{3}{2}u^5\right)\log(1+u)+\left(4-\frac{15}{2}u+5u^3-\frac{3}{2}u^5\right)\log(1-u)+29u^2-3u^4-40u^2\log2.\\
\nonumber
F_{2}^{1}(u)&=&\left(4-20u^2-20u^3+4u^5\right)\log(u+1)+\left(20u^2-20u^3+4u^5-4\right)\log(u-1)+22u+4u^3+(40u^3-8u^5)\log u.
\\
\nonumber
F_{1}^{2}(u)&=&\left(\frac{15}{8}-\frac{6}{7 u}+12 u -\frac{195}{8} u^2 +\frac{129}{8} u^4 -\frac{267}{56} u^6\right)\log \left(1-u\right)-\frac{311 u}{28}+\frac{1457 u^3}{14}-\frac{267 u^5}{28}\\
\nonumber &&\:-
\left(\frac{15}{8}+\frac{6}{7 u}-12 u-\frac{195}{8} u^2+\frac{129}{8} u^4-\frac{267}{56} u^6\right)\log(1+u)+\left(8u-140 u^3\right)\log 2\\
\nonumber 
F_{2}^{2}(u)&=&-\frac{58}{7}+\frac{550 u^2}{7}+\frac{92 u^4}{7}+\left(\frac{6}{7 u}-16 u +70 u^3 -68 u^4 +\frac{92}{7}u^6\right)\log (u-1)\\
\nonumber 
&&\:+\left(136 u^4 -\frac{184}{7} u^6\right)\log u+\left(-\frac{6}{7 u}+16 u -70 u^3-68 u^4+\frac{92}{7} u^6\right)\log (u+1) \\
\nonumber
F_{1}^{3}(u)&=&\left(\frac{15}{4}-\frac{12}{7 u}+16 u -\frac{135}{4} u^2 +\frac{89}{4} u^4 -\frac{183}{28} u^6\right)\log(1-u)-\frac{311 u}{14}+\frac{1051 u^3}{7}-\frac{183 u^5}{14}\\
\nonumber 
&&\:-\left(\frac{15}{4}+\frac{12}{7 u}-16 u-\frac{135}{4} u^2 +\frac{89}{4} u^4-\frac{183}{28} u^6\right)\log(1+u)+\left(16 u-200 u^3\right)\log 2\\
\nonumber
F_{2}^{3}(u)&=&-\frac{116}{7}+\frac{792 u^2}{7}+\frac{128 u^4}{7}+\left(\frac{12}{7 u}-24 u+100 u^3-96 u^4+\frac{128}{7} u^6\right)\log (u-1)\\
\nonumber 
&&\:+\left(192 u^4 -\frac{256}{7} u^6\right)\log u+\left(-\frac{12}{7 u}+24 u -100 u^3-96 u^4+\frac{128}{7} u^6\right)\log (u+1)
\\
\nonumber
F_{1}^{4}(u)&=&\left(\frac{15}{4}-\frac{12}{7 u}+4 u -\frac{45}{4} u^2+\frac{29}{4} u^4 -\frac{57}{28} u^6\right)\log (1-u)-\frac{311 u}{14}+\frac{442 u^3}{7}-\frac{57 u^5}{14}+\left(16 u-80 u^3\right)\log 2\\
\nonumber 
&&\:+
\left(-\frac{15}{4}-\frac{12}{7 u}+4 u +\frac{45}{4} u^2-\frac{29}{4} u^4 +\frac{57}{28} u^6\right)\log (1+u)\\
\nonumber 
F_{2}^{4}(u)&=&-\frac{116}{7}+\frac{330 u^2}{7}+\frac{44 u^4}{7}+\left(\frac{12}{7 u}-12 u+40 u^3-36 u^4+\frac{44}{7} u^6\right)\log(u-1)\\
\nonumber
&&\:+
\left(72 u^4-\frac{88}{7} u^6\right)\log u+\left(-\frac{12}{7 u}+12 u -40 u^3 -36 u^4 +\frac{44}{7} u^6\right)\log (u+1)\\
\nonumber
F_{1}^{5}(u)&=&\frac{311 u}{420}-\frac{239 u^3}{210}+\frac{u^5}{28}-\left[\frac{8}{15}u -\frac{4}{3}u^3\right]\log 2-\left[\frac{1}{8}-\frac{2}{35 u}-\frac{1}{8}u^2+\frac{3}{40}u^4-\frac{1}{56}u^6\right]\log[1-u]\\
\nonumber
&&\:+\left[\frac{1}{8}+\frac{2}{35 u}-\frac{1}{8}u^2+\frac{3}{40}u^4-\frac{1}{56}u^6\right]\log[1+u]\\
\nonumber
F_{2}^{5}(u)&=&\frac{58}{105}-\frac{88 u^2}{105}-\frac{8 u^4}{105}-\left[\frac{16}{15}u^4-\frac{16}{105}u^6\right]\log u-\left[\frac{2}{35 u}-\frac{4}{15}u+\frac{2}{3}u^3-\frac{8}{15}u^4+\frac{8}{105}u^6\right]\log[u-1]\\
\nonumber
&&\:+\left[\frac{2}{35 u}-\frac{4}{15}u+\frac{2}{3}u^3+\frac{8}{15}u^4-\frac{8}{105}u^6\right]\log[u+1]
\end{eqnarray}

\newpage

\section{APPENDIX B}
The expressions of the functions $F^{abc}_1(u)$, $F^{abc}_2(u)$ and $F^{abc}_3(u)$ appearing in Eq. (19) are
\begin{eqnarray*}
F_1^{abc}(u)&=&\frac{u^2}{2}\left(11b^4c+18ab^3c^2+18a^2b^2c^3+11a^3bc^4\right)+u^3(18b^3c-9b^4c+6ab^2c^2
-9b^3c^2-3ab^3c^2+18a^2bc^3\\&&\:-
3ab^2c^3-9a^2b^2c^3-9a^2bc^4)+u^4(6b^2c-6b^3c+2b^4c+6abc^2-6b^2c^2-6ab^2c^2+3b^3c^2-\frac{1}{2}ab^3c^2\\&&\:-6abc^3-\frac{1}{2}b^2c^3+3ab^2c^3+2abc^4)+u^5(4bc-6b^2c-b^3c+\frac{3}{2}b^4c-6bc^2+6b^2c^2+\frac{1}{2}b^3c^2-bc^3\\&&\:+\frac{1}{2}b^2c^3+\frac{3}{2}bc^4)\\&&\:
+\log 2 \Bigg\{-20u^2(b^4c+a^3bc^4)+u^3(30b^4c-60b^3c+30b^3c^2-60a^2bc^3+30a^2b^2c^3\\&&\:+
30a^2bc^4)+ u^5(80bc-120b^2c+80b^3c-20b^4c-120bc^2+120b^2c^2-40b^3c^2+80bc^3-40b^2c^3\\&&\:-
20bc^4)\Bigg\}+10u^5bc(2-b-c)^3\log(2(2-b-c)u)\\&&\:+
\Bigg\{\frac{1}{2}(b^5-5a^2b^3c^2-5a^3b^2c^3+a^5c^5)+\frac{15u}{4}(b^4-b^5-2a^2b^2c^2+2a^2b^3c^2+a^4c^4-a^4bc^4)\\&&\:+
10u^2(b^3-2b^4+b^5+a^3c^3-2a^3bc^3+a^3b^2c^3)+10u^3(b^2-3b^3+3b^4-b^5+a^2c^2-3a^2bc^2+3a^2b^2c^2\\&&\:
-a^2b^3c^2)+
u^5(-4+20b-40b^2+40b^3-20b^4+4b^5)\Bigg\}\log(b+ac+2u-2bu)\\&&\:+
\Bigg\{(\frac{15}{4}a^4bc^4u+10a^3u^2(2bc^3-b^2c^3)+
a^2u^3(30bc^2-30b^2c^2+\frac{15}{2}b^3c^2)\\&&\:+u^5(-20b+40b^2-30b^3+10b^4-\frac{5}{4}b^5)\Bigg\}\log(2ac+4u-2bu)\\&&\:
+\Bigg\{\frac{1}{2}(b^5-5a^2b^3c^2-5a^3b^2c^3+a^5c^5)+\frac{15u}{4}(b^4-b^4c-2a^2b^2c^2+2a^2b^2c^3+a^4c^4-a^4c^5)\\&&\:+
10u^2(b^3-2b^3c+b^3c^2+a^3c^3-2a^3c^4+a^3c^5)+10u^3(b^2-3b^2c+a^2c^2+3b^2c^2-3a^2c^3-b^2c^3\\&&\:+
3a^2c^4-a^2c^5)+u^5(-4+20c-40c^2+40c^3-20c^4+4c^5)\Bigg\}\log(b+ac+2u-2cu)\\&&\:+
\Bigg\{\frac{15}{4}b^4cu+u^2(20b^3c-10b^3c^2)+u^3(30b^2c-30b^2c^2+\frac{15}{2}b^2c^3)+u^5(40c^2-20c-30c^3+10c^4\\&&\:-
\frac{5}{4}c^5)\Bigg\}\log(2b+4u-2cu)\\&&\:+
\Bigg\{\frac{1}{2}(-b^5+5a^2b^3c^2+5a^3b^2c^3-a^5c^5)+\frac{15u}{4}(-b^4+b^5+b^4c+2a^2b^2c^2-2a^2b^3c^2-2a^2b^2c^3\\&&\:-
a^4c^4+a^4bc^4+a^4c^5)
+u^2(-10b^3+20b^4-10b^5+20b^3c-20b^4c-10b^3c^2-10a^3c^3+20a^3bc^3\\&&\:-
10a^3b^2c^3+20a^3c^4-20a^3bc^4-10a^3c^5)\\&&\:+
u^3(-10b^2+30b^3-30b^4+10b^5+30b^2c-60b^3c+30b^4c-10a^2c^2+30a^2bc^2-30b^2c^2-30a^2b^2c^2\\&&\:+
30b^3c^2+10a^2b^3c^2+30a^2c^3-60a^2bc^3+10b^2c^3+30a^2b^2c^3-30a^2c^4+30a^2bc^4+10a^2c^5)\\&&\:+
u^5(4-20b+40b^2-40b^3+20b^4-4b^5-20c+80bc-120b^2c+80b^3c-20b^4c+40c^2-120bc^2\\&&\:+ 
120b^2c^2-40b^3c^2-40c^3+80bc^3-40b^2c^3+20c^4-20bc^4-4c^5)\Bigg\}\log(b+ac+2u-2bu-2cu)
\end{eqnarray*}
\begin{eqnarray*}
&&\:+
\Bigg\{-\frac{15}{4}a^4bc^4u+u^2(-20a^3bc^3+10a^3b^2c^3+20a^3bc^4)+u^3(-30a^2bc^2+30a^2b^2c^2-\frac{15}{2}a^2b^3c^2+60a^2bc^3\\&&\:-
30a^2b^2c^3-30a^2bc^4)+u^5(20b-40b^2+30b^3-10b^4+\frac{5}{4}b^5-80bc+120b^2c-60b^3c+10b^4c+120bc^2\\&&\:-
120b^2c^2+30b^3c^2-80bc^3+40b^2c^3+20bc^4)\Bigg\}\log(2ac+4u-2bu-4cu)\\&&\:+
\Bigg\{-\frac{15}{4}b^4cu+u^2(-20b^3c+20b^4c+10b^3c^2)+u^3(-30b^2c+60b^3c-30b^4c+30b^2c^2-30b^3c^2-\frac{15}{2}b^2c^3)\\&&\:
+u^5(20c-80bc+120b^2c-80b^3c+20b^4c-40c^2+120bc^2-120b^2c^2+40b^3c^2+30c^3-60bc^3+30b^2c^3-10c^4\\&&\:+ 10bc^4+\frac{5}{4}c^5)\Bigg\}\log(4u-2cu+2b-4bu)\\&&\:-
\Bigg\{\frac{1}{2}(b^5-5a^2b^3c^2-5a^3b^2c^3+a^5c^5)+\frac{15u}{4}(b^4-2a^2b^2c^2+a^4c^4)+10u^2(b^3+a^3c^3)\\&&\:+
10u^3(b^2+a^2c^2)-4u^5\Bigg\}\log(b+ac+2u)\\
F_2^{abc}(u)&=&\{\frac{1}{2}(b^5-5a^2b^3c^2-5a^3b^2c^3+a^5c^5)+\frac{15u}{4}(-b^4+2a^2b^2c^2-a^4c^4)\\&&\:+ 10u^2(b^3+a^3c^3)-10u^3(b^2+a^2c^2)+4u^5\}\log(2u-ac-b)\\
&&\:+\{\frac{1}{2}(b^5-5a^2b^3c^2+5a^3b^2c^3-a^5c^5)+\frac{15u}{4}(b^4-2a^2b^2c^2+a^4c^4)\\&&\:+ 10u^2(b^3-a^3c^3)+10u^3(b^2+a^2c^2)-4u^5\}\log(2u-ac+b)\\
&&\:+\{\frac{1}{2}(-b^5+5a^2b^3c^2-5a^3b^2c^3+a^5c^5)+\frac{15u}{4}(b^4-2a^2b^2c^2+a^4c^4)\\&&\:+ 10u^2(-b^3+a^3c^3)+10u^3(b^2+a^2c^2)-4u^5\}\log(2u+ac-b)\\
&&\:-\{\frac{1}{2}(b^5-5a^2b^3c^2-5a^3b^2c^3+a^5c^5)+\frac{15u}{4}(b^4-2a^2b^2c^2+a^4c^4)\\&&\:+ 10u^2(b^3+a^3c^3)+10u^3(b^2+a^2c^2)-4u^5\}\log(2u+ac+b)\\
&&\:+ 11abcu(b^2+a^2c^2)+4abcu^3
\end{eqnarray*}
\begin{eqnarray*}
F_3^{abc}(u)&=&\frac{11u}{2}\left(ab^4c+a^3b^2c^3\right)+u^2\left(18ab^3c-9ab^4c+11a^3bc^3-\frac{11}{2}a^3b^2c^3\right)+u^3\left(6ab^2c-6ab^3c +2ab^4c\right)\\&&\:+
u^4\left(4abc-6ab^2c-ab^3c+\frac{3}{2}ab^4c\right)+u^2\left(-40a^3bc^3+20a^3b^2c^3\right)\log 2\\&&\:
+\Bigg\{-\frac{1}{2} b^5 +\frac{5}{2} a^2 b^3 c^2 -\frac{5}{2} a^3 b^2 c^3 +\frac{1}{2} a^5 c^5 +u(-\frac{15}{4} b^4 +\frac{15}{4} b^5 +\frac{15}{2} a^2 b^2 c^2 -\frac{15}{2} a^2 b^3 c^2 -\frac{15}{4} a^4 c^4\\&&\:
+\frac{15}{4} a^4 b c^4) +u^2(-10 b^3 +20 b^4 -10 b^5 +10 a^3 c^3 -20 a^3 b c^3 +10 a^3 b^2 c^3)\\&&\:
+u^3(-10 b^2 +30 b^3 -30 b^4 +10 b^5 -10 a^2 c^2 +30 a^2 b c^2 -30 a^2 b^2 c^2 +10 a^2 b^3 c^2) \\&&\:
+u^5(4 -20 b +40 b^2 -40 b^3 +20 b^4  -4 b^5)\Bigg\}\log(b-ac+2u-2bu)\\&&\:
+\Bigg\{\frac{1}{2} b^5 -\frac{5}{2} a^2 b^3 c^2 -\frac{5}{2} a^3 b^2 c^3 +\frac{1}{2} a^5 c^5+u(\frac{15}{4} b^4 -\frac{15}{4} b^5 -\frac{15}{2} a^2 b^2 c^2 +\frac{15}{2} a^2 b^3 c^2 +\frac{15}{4} a^4 c^4 -\frac{15}{4} a^4 b c^4)\\&&\:
+u^2(10 b^3 -20 b^4 +10 b^5 +10 a^3 c^3  -20 a^3 b c^3 +10 a^3 b^2 c^3)\\&&\:
+u^3(10 b^2 -30 b^3 +30 b^4 -10 b^5 +10 a^2 c^2 -30 a^2 b c^2+30 a^2 b^2 c^2  -10 a^2 b^3 c^2)\\&&\:
+u^5(-4 +20 b -40 b^2  +40 b^3 -20 b^4 +4 b^5\Bigg\}\log(b+ac+2u-2bu) \\&&\:
+\Bigg\{\frac{15}{4} a^4 b c^4 u +u^2(20 a^3 b c^3 -10 a^3 b^2 c^3 )+u^3(30 a^2 b c^2 -30 a^2 b^2 c^2 +\frac{15}{2}a^2b^3c^2)\\&&\:
+u^5(-20b+40b^2-30b^3+10b^4-\frac{5}{4}b^5)\Bigg\}\log(2ac+4u-2bu)\\&&\:
-\Bigg\{\frac{b^5}{2}-\frac{5}{2}a^2b^3c^2-\frac{5}{2}a^3b^2c^3+\frac{a^5c^5}{2}+u(\frac{15b^4}{4}-\frac{15}{2}a^2b^2c^2+\frac{15}{4}a^4c^4)+u^2(10b^3+10a^3c^3)\\&&\:
+u^3(10b^2+10a^2c^2)-4u^5\Bigg\}\log(b+ac+2u)\\&&\:
+\frac{1}{8}(b-ac+2u)^3(4b^2+4a^2c^2-6acu-4u^2+3b(4ac+2u))\log(b-ac+2u)\\&&\:
+\frac{5}{64}\left[-6ac-4u+2bu\right]\left[2ac-4u+2bu\right]^3\log(4u-2bu-2ac)
\end{eqnarray*}


\begin{references}
\bibitem{bernard} V. Bernard and Nguyen Van Giai, Nucl. Phys. A 348, 75 (1980). 
\bibitem{colo} G. Col\`o, H. Sagawa, and P.F. Bortignon, Phys. Rev. C 82, 064307 (2010).
\bibitem{ring} E. Litvinova, P. Ring, and V. Tselyaev, Phys. Rev. C 75, 064308 (2007).
\bibitem{huang} K. Huang, {\it Statistical Mechanics} (Wiley, New York, (1987)).
\bibitem{bulgac} A. Bulgac and Y. Yu, Phys. Rev. Lett. 88, 042504 (2002).
\bibitem{bruun} G. Bruun, Y. Castin, R. Dum, and K. Burnett, Eur. Phys. J. D 7, 433 (1999).
\bibitem{grasso} M. Grasso and M. Urban, Phys. Rev. A 68, 033610 (2003).
\bibitem{dr1} Gerard't Hooft and M.J.G. Veltman, Nucl Phys. B 44, 189 (1972).
\bibitem{dr2} Gerard't Hooft, Nucl Phys. B 61, 455 (1973).
\bibitem{dr3} C.G. Bollini and J.J. Giambiagi, Nuovo Cimento B 12, 20 (1972).
\bibitem{leib} George Leibbrandt, Rev. Mod. Phys. 47, 849 (1975).
\bibitem{mog} K. Moghrabi, M. Grasso, G. Col\`o, and N. Van Giai, Phys. Rev. Lett. 105, 262501 (2010).
\bibitem{skp} J. Dobaczewski, H. Flocard, and J. Treiner, Nucl. Phys. A 422, 103 (1984). 
\bibitem{sly5} E. Chabanat, P. Bonche, P. Haensel, J. Meyer, and R. Schaeffer, Nucl. Phys. A 627, 710 (1997); {\it ibid.} A 635, 231 (1998); {\it ibid.} A 643, 441 (1998). 
\bibitem{meyer} J. Meyer, Ann. Phys. Fr. 28, n. 3 (2003).
\bibitem{gambacurta} D. Gambacurta, M. Grasso, and F. Catara, Phys. Rev. C 81, 054312 (2010).
\bibitem{pillet} N. Pillet, J.-F. Berger, and E. Caurier, Phys. Rev. C 78, 024305 (2008).
\bibitem{bevington} P. R. Bevington and D. K. Robinson, {\it Data reduction and error analysis for physical sciences}, Second Edition (McGraw-Hill, NeW York 1992).

\end{references}
\end{document}